\documentclass[12pt]{spieman}  
\usepackage{amsmath,amsfonts,amssymb}
\usepackage{graphicx}
\usepackage{setspace}
\usepackage{tocloft}
\usepackage{subcaption}
\usepackage{xcolor}

\usepackage{etoolbox}

\title{Underdetermined Polarimetric Measurements for Mueller Extrapolations}

\author[a*]{Quinn Jarecki}
\author[a]{Meredith Kupinski}
\affil[a]{Wyant College of Optical Sciences, University of Arizona, Tucson, USA}

\cftpagenumbersoff{figure}
\cftpagenumbersoff{table} 
\begin{document} 
\maketitle

\begin{abstract}
 Polarized light-matter interactions are mathematically described by the Mueller matrix (MM)-valued polarized bidirectional reflectance distribution function (pBRDF). A pBRDF is parameterized by 16 degrees of freedom that depend upon scattering geometry. A triple degenerate (TD) MM assumption reduces the degrees of freedom to eight: one for reflectance, six for non-depolarizing properties, and one for depolarization. When the non-depolarizing dominant process is known or assumed (\emph{e.g.} Fresnel reflection), the degrees of freedom are further reduced to two. 
 For a given material, if the TD model is appropriate and the dominant non-depolarizing process is known, then these two degrees of freedom can be estimated from as few as two polarimetric measurements. Thus, the MM can be extrapolated from a reduced number of measurements. The primary contribution of this work is the development and demonstration of a linear estimator for a MM's dominant eigenvalue (\emph{i.e.} single depolarization parameter) which requires fewer measurements than a full MM reconstruction. 
 MM extrapolations from single snapshot acquisitions with a Sony Triton 5.0MP Polarization Camera are performed at 30 acquisition geometries and two wavelengths on an ensemble of LEGO bricks treated to have varying surface roughness. 
 These extrapolated MMs are compared to MMs reconstructed from a complete dual rotating retarder (DRR) Mueller imaging polarimeter. 
The flux error mean and mode are 11.06\% and 1.03\%, respectively,
despite a 10$\times$ reduction in the number of polarimetric measurements.
\end{abstract}

\keywords{Polarization, depolarization, Mueller matrix extrapolation, pBRDF, linear Stokes camera, snapshot polarimeter, partial polarimetry}

{\noindent \footnotesize\textbf{*}Quinn Jarecki,  \linkable{qjarecki@email.arizona.edu} }

\begin{spacing}{2} 

\section{Introduction}
\label{sect:intro}

Polarized light-matter interactions are mathematically described by Mueller matrix (MM)-valued functions called polarized bidirectional reflectance distribution functions (pBRDF) corresponding to 16 degrees of freedom at each scattering geometry. While analytical models exist, building an experimental model with no prior information requires a minimum of 16 polarimetric measurements at each combination of input and output scattering geometries. More than 16 polarimetric measurements creates an overdetermined system and a pseudoinverse  solution that becomes more robust to noise as more linearly dependent measurements are taken\cite{Smith:02, Twietmeyer:08}. 


Cloude coherency matrix eigenanalysis is a standard technique for analyzing MMs\cite{cloude_eigenspectrum,cloude_entropy, Cloude1986GroupTA,cloudepottier,Cloude1994scatterers,Kupinski:18,Sheppard:16,Sheppard:18}. A MM with three equal coherency eigenvalues, a triple degeneracy (TD), has only eight degrees of freedom\cite{Li_singleParam}. When the dominant non-depolarizing process is known, this reduces further to two degrees of freedom. These degrees of freedom correspond to the average reflectance and the dominant coherency eigenvalue. Based on empirical observation, the non-depolarizing process is assumed to be Fresnel reflection, even in the case where subsurface scattering is stronger. 

The objective of this work is to use prior knowledge about a material to extrapolate its MM from a small quantity of measurements. 
Our strategy is to relate polarimetric measurements to the two degrees of a freedom of a TD MM model when the dominant process is known, rather than to relate measurements to the MM directly. 
The primary contribution of this work is a linear estimator for a MM's dominant coherency eigenvalue which requires as few as two polarimetric measurements. This is the first method known to the authors for extrapolating depolarizing MMs with rank four coherency matrices from fewer than 16 measurements. We make use of a single illumination polarization state and a commercial division of focal plane (DoFP) linear Stokes camera, meaning that four measurements can be performed in a single snapshot acquisition. Experimental results using measurements taken in a snapshot configuration with a Sony Triton 5.0MP Polarization Camera are presented and compared to a complete dual rotating retarder (DRR) Mueller polarimeter.

\section{Background}

The bidirectional reflectance distribution function (BRDF) of a material is a radiometric property defined as the ratio of differential output radiance to differential input irradiance\cite{Bartell_BRDF_theory}.
The term "bidirectional" refers to the dependence on both the incident and observation directions. These are specified with the angles $\theta_i$ and $\phi_i$ which are the zenith and azimuth angles of the incident direction, respectively, and $\theta_o$ and $\phi_o$  which are the zenith and azimuth angles of the observation direction, respectively. $\theta_i$ and $\theta_o$ are also referred to as the incident and scattered angles, respectively. 

Torrance and Sparrow introduced the theory of microfacets to explain off-specular scattering observed in most BRDFs\cite{Torrance:67}.
According to the theory, off-specular reflection from a surface with normal $\widehat{\mathbf{n}}$ is the result of small, randomly-oriented mirror-like microfacets with varying surface normals $\widehat{\mathbf{h}}$ combined with a diffuse component. 
The light which impinges on a microfacet obeys the law of reflection, so its behavior is angle-dependent. For a given pair of incident and exitant directions, $\widehat{\boldsymbol{\omega}}_i$ and $\widehat{\boldsymbol{\omega}}_o$ respectively, the amount of reflected light depends on the average of the Fresnel reflectance coefficients, the probability of the microfacet normal satisfying the law of reflection, and the probability of interaction between adjacent microfacets. Light incident on one microfacet may be blocked by a neighbor preventing full illumination (called shadowing) or light reflected from a microfacet may be blocked by a neighbor from reaching the observer (called masking). The diffuse component is an angle-independent term that describes both light scattered by multiple microfacets before reaching the observer and light that transmits into and then out of the material. The microfacet BRDF model is characterized by the angle-dependent distribution on microfacet orientations, shadowing-masking functions, and the relative weight of specular to diffuse. 

Whereas a scalar BRDF is a single-valued function of four variables, polarized bidirectional reflectance distribution functions (pBRDF) are Mueller matrix-valued functions of four variables. A Mueller matrix (MM) is a 16-element matrix that describes the linear transformation of a polarization state due to a light-matter interaction,
\begin{equation}
    \mathbf{M}=M_{00}\mathbf{m}=M_{00}\begin{pmatrix}
       1   & m_{01} & m_{02} & m_{03}\\
    m_{10} & m_{11} & m_{12} & m_{13}\\
    m_{20} & m_{21} & m_{22} & m_{23}\\
    m_{30} & m_{31} & m_{32} & m_{33}
    \end{pmatrix}.
    \label{eq:MM}
\end{equation}
Having 16 elements means there are 16 degrees of freedom: one for average reflectance, three for diattenuation, three for retardance, and nine for depolarization\cite{chipman}. Unfortunately, the retardance and depolarization degrees of freedom do not correspond to individual Mueller elements and are instead coupled among multiple Mueller elements. Depolarization is the random modulation of polarization in time, space, and/or wavelength faster than the particular detector in use can resolve\cite{Billings_monochromatic}. The average reflectance is $M_{00}$, which can be factored out to produce a normalized MM that describes polarimetry separately from radiometry. In this work, a MM designated with a lowercase letter is normalized as in Eq.~(\ref{eq:MM}).
A challenge inherent to pBRDFs is that polarization properties are defined in the plane transverse to the ray direction. When considering all possible combinations of input and output directions, keeping track of the basis vectors describing the transverse planes becomes paramount. 

One of the first models for a pBRDF is an extension of the popular microfacet model by Priest and Germer in 2000 \cite{priest_germer}. In this model, the Fresnel equations are not combined to an average term and, instead, the Fresnel reflection Mueller matrix is used with careful consideration for the coordinate systems before and after scattering. This Mueller matrix has the form
\begin{equation}
    \mathbf{M}(\widehat{\boldsymbol{\omega}}_i,\widehat{\boldsymbol{\omega}}_o,n_0,n_1)=\mathbf{R}(\alpha_o)\mathbf{F}_R(n_0,n_1,\theta_d)\mathbf{R}(-\alpha_i)
    \label{eq:fresnRot}
\end{equation}
where the $\mathbf{R}(\alpha)$ are Mueller rotation matrices of the form
\begin{equation}
    \mathbf{R}(\alpha)=\begin{pmatrix}
    1&0&0&0\\
    0&\cos(2\alpha)&-\sin(2\alpha)&0\\
    0&\sin(2\alpha)&\cos(2\alpha)&0\\
    0&0&0&1
    \end{pmatrix},
\end{equation}
$\alpha_i$ and $\alpha_o$ are the rotations from the polarization state generator (PSG) coordinates to local microfacet coordinates and from local microfacet coordinates to the polarization state analyzer (PSA) coordinates, respectively, and $\mathbf{F}_R(n_0,n_1,\theta_d)$ is the Fresnel reflection Mueller matrix of the form
\begin{equation}
    \mathbf{F}_R(n_0,n_1,\theta_d)=\frac{1}{2}\begin{pmatrix}
    |r_p|^2+|r_s|^2 & |r_p|^2-|r_s|^2 & 0 &0\\
    |r_p|^2-|r_s|^2 & |r_p|^2+|r_s|^2 & 0 & 0\\
    0& 0 &2\mathrm{Re}(r_p^*r_s)&2\mathrm{Im}(r_p^*r_s)\\
    0 & 0 & -2\mathrm{Im}(r_p^*r_s) & 2\mathrm{Re}(r_p^*r_s)
    \end{pmatrix}
    \label{eq:Fresnel}
\end{equation}
where $r_s$ and $r_p$ are the well known Fresnel amplitude reflection coefficients that depend on the refractive indices $n_0$ and $n_1$ and the angle of incidence onto the microfacet $\theta_d$. It is worth noting that typically $\alpha_i$ is considered to be the rotation from the coordinate system defined by $\widehat{\mathbf{n}}$ and $\widehat{\boldsymbol\omega}_i$ to local microfacet coordinates, and $\alpha_o$ is the rotation from the microfacet coordinates to the coordinate system defined by $\widehat{\mathbf{n}}$ and $\widehat{\boldsymbol\omega}_o$. These rotations are needed for the forward problem of computer graphics rendering from a pBRDF, but since the goal of this work is to recover MMs, the input and output coordinate systems are different. 

Other similar pBRDFs based on the microfacet model make use of some form of depolarizing component. The most basic approach is to set the diffuse term to an ideal depolarizer. The ideal depolarizer MM has the form
\begin{equation}
    \mathbf{m}_{ID}=\begin{pmatrix}
    1&0&0&0\\
    0&0&0&0\\
    0&0&0&0\\
    0&0&0&0
    \end{pmatrix},
\end{equation}
and has the property of converting any incident polarization state to unpolarized light. 

Another pBRDF which interprets the diffuse term as a depolarizer is proposed by Baek et al. in 2018 \cite{Baek}. In this model, the ideal depolarizer is sandwiched between two Fresnel transmission matrices. The Fresnel transmission matrix has the a similar form as in Eq.~(\ref{eq:Fresnel}), but with the amplitude transmission coefficients replacing the reflection coefficients. The rationale for this diffuse term is that some fraction of the incident light undergoes Fresnel transmission into the material, undergoes subsurface scattering which completely depolarizes the light, and then is transmitted out. A pBRDF model by Kondo et al. is a sum of a specular and two diffuse terms \cite{kondo2020accurate}. 

As stated above, a Mueller matrix has 16 degrees of freedom and therefore must be constrained by at least 16 linearly independent measurements. The most common architecture for a complete MM polarimeter is a DRR polarimeter which takes a sequence of measurements with different retarder positions\cite{Azzam_1978,Goldstein_1992}. Taking 16 or more sequential measurements at sufficiently fine scattering geometry sampling to build an empirical pBRDF is a time-intensive process. This motivates the development of snapshot polarimeters. Technologies for snapshot acquisition of MMs include encoding polarization to wavelength \cite{Hagen_Oka_Dereniak_2007,Dubreuil:07,Lemaillet:08,Alenen_Tyo_2012}, channeled polarimetry using polarization gratings\cite{Kudenov:12,Kudenov_Mallik_Escuti_Hagen_Oka_Dereniak_2013}, structured illumination combined with a DoFP full Stokes camera\cite{Zaidi_McEldowney_Lee_Chao_Lu_2022}, and splitting the beam to enable multiple DoFP cameras\cite{Huang:21}. Hybrid approaches also exist which use rotating retarders for illumination but decrease the number of acquisitions by using a DoFP camera\cite{Vaughn_Rodriguez-Herrera_2015}.

These technologies are fast, but remain complex. Using partial polarimetry simplifies the measurement requirements. Partial polarimetric systems take fewer than 16 linearly independent measurements, making them underdetermined and incapable of reconstructing the full MM. 
Recent work by Gonzalez et al. demonstrated analysis and decomposition of 3$\times$4 partial MMs measured using four polarized illumination states and a snapshot linear Stokes camera\cite{Gonzalez_2021}. 
A partial polarimeter which only makes use of linear illumination and analyzer states can at most reconstruct the upper 3$\times$3 portion of a MM. In 2013 Swami et al. showed that, for a non-depolarizing MM, symmetry arguments can be applied to the linear partial MM to obtain the full 4$\times$4 matrix\cite{SWAMI201318}. In 2019, Ossikovski and Arteaga showed symmetry arguments for obtaining a full 4$\times$4 MM from 12 elements where a row or column is missing\cite{Arteaga_12element} or from nine elements where a row and column are missing\cite{Ossikovski_9element}. In the 12-element case, it is possible to recover a depolarizing MM but only if it obeys certain symmetry constraints and has only two non-zero coherency eigenvalues.

The pBRDF model explored in this work was derived by Li and Kupinski in 2021\cite{Li_singleParam} and demonstrated in polarimetric computer graphics renderings by Omer and Kupinski in 2022\cite{KhalidRendering}. 
This model is based on the coherency matrix introduced by Cloude\cite{Cloude1986GroupTA,cloudepottier}. The coherency matrix and its eigendecomposition (sometimes called the spectral decomposition) are standard techniques for analyzing MMs \cite{Cloude1994scatterers,Kupinski:18,Sheppard:16,Sheppard:18,cloude_eigenspectrum,cloude_entropy, Cloude1986GroupTA,cloudepottier}.
Using the spectral decomposition, a depolarizing MM is rewritten as a convex sum of up to four non-depolarizing MMs,
\begin{eqnarray}
    \mathbf{m}=\sum_{n=0}^3\xi_n\widehat{\mathbf{m}}_n,
    \label{eq:cloudeDecomp}
\end{eqnarray}
where $\xi_n$ are the Cloude coherency matrix eigenvalues normalized so that $\sum_{n=0}^3\xi_n=1$ and $\widehat{\mathbf{m}}_n$ are the non-depolarizing (indicated with the hat $\widehat{\cdot}$ ) MMs which also have the property that $\frac{1}{4}\sum_{n=0}^3\widehat{\mathbf{m}}_n=\mathbf{m}_{ID}$. Li and Kupinski showed that when the smaller three eigenvalues of the Cloude coherency matrix are equal, the MM has the form
\begin{equation}
    \mathbf{M}=\frac{4M_{00}}{3}\left[\left(\xi_0-\frac{1}{4}\right)\widehat{\mathbf{m}}_0+\left(1-\xi_0\right)\mathbf{m}_{ID}\right],
    \label{eq:1par_a}
\end{equation}
where $\widehat{\mathbf{m}}_0$ is the dominant non-depolarizing MM and $\xi_0$ now controls the relative weight between this dominant process and the ideal depolarizer. This MM is referred to as being triply degenerate (TD) because the last three eigenvalues are identical. The 16 degrees of freedom of a general MM are reduced to eight: one for $M_{00}$, one for $\xi_0$, three for the diattenuation orientation and magnitude of $\widehat{\mathbf{m}}_0$, and three for the retardance orientation and magnitude of $\widehat{\mathbf{m}}_0$. Both $\widehat{\mathbf{m}}_0$ and $\xi_0$ are functions of the scattering geometry: $\theta_i$, $\phi_i$, $\theta_o$, and $\phi_o$, though this dependence is omitted for brevity. The decomposition of a MM into a non-depolarizing component and an ideal depolarizer has appeared in the literature before, such as in the textbook by Brosseau where a MM is written as a sum of a non-depolarizing MM which depends on the input Stokes vector and an ideal depolarizer\cite{Brosseau_1998}. For the TD model, the non-depolarizing MM does not depend on the input Stokes vector and the relative contribution of each component is determined by the unique coherency eigenvalue.

When the dominant process is believed to be Fresnel reflection, the angular dependence of $\widehat{\mathbf{m}}_0$ is known and the degrees of freedom are reduced to two.
Based on knowing \emph{a priori} the dominant coherent process $\widehat{\mathbf{m}}_0$, Li and Kupinski proposed a method to measure $\xi_0$ using only two linearly independent polarimetric measurements\cite{Li_singleParam}. The method presented in our work would, in principle, work with only two linearly independent polarimetric measurements as well. However with a commercially available linear Stokes camera, such as the Sony Triton 5.0MP Polarization Camera used in this work, simultaneous acquisition of four (three linearly independent) polarimetric measurements is possible. The MM extrapolations presented make use of all four measurements. Jarecki and Kupinski presented initial results for low albedo measurements in 2022\cite{quinn2022}. In this work, high albedo measurements and additional analysis are included.

\section{Methods}
\subsection{Polarimetry}
The MM of a particular light-matter interaction is measured by illuminating the sample with a known polarization state generator (PSG) then measuring the scattered light through a known polarization state analyzer (PSA) as in
\begin{equation}
    p_j=\mathbf{a}_j^{\mathrm{T}}\mathbf{M}\mathbf{g}_j=\mathbf{w}_j\mathbf{M}_{16\times1},
    \label{eq:flux1}
\end{equation}
where $\mathbf{M}$ is the object MM, $\mathbf{g}_j$ is the $j^{th}$ PSG state, $\mathbf{a}_j$ is the $j^{th}$ PSA state, and $p_j$ is the $j^{th}$ irradiance measurement at the detector. This measurement equation is rewritten in the right-hand side of Eq.~(\ref{eq:flux1}) as an inner product of two vectors, where $\mathbf{w}_j=\mathbf{a}_j\otimes\mathbf{g}_j^{\mathrm{T}}$ and $\mathbf{M}_{16\times1}$ is the object MM elements ordered in a vector form. Conventional full polarimetry requires that at least $J=16$ linearly-independent measurements be performed with different illumination and analyzer states to constrain the 16 degrees of freedom of $\mathbf{M}$. Most polarimeters have $J>16$ to create an overdetermined system for mitigating the effects of noise\cite{Smith:02,Twietmeyer:08}. The MM estimate is then reconstructed via the pseudoinverse as in
\begin{equation}
     \widetilde{\mathbf{M}}_{16\times1}=\mathbf{W}^{+}\mathbf{P},
     \label{eq:MM_recon}
\end{equation}
where $\mathbf{W}$ is called the polarimetric measurement matrix and has rows $\mathbf{w}_j$, $\mathbf{P}$ is the vector of flux measurements with elements $p_j$, and the tilde on $\widetilde{\mathbf{M}}$ indicates the quantity is an estimate.


\subsection{Triple Degenerate Model}
Using the Cloude coherency eigenanalysis of MMs, Li and Kupinski derived the triple degenerate (TD) model where a MM consists of a weighted sum of a dominant non-depolarizing MM and an ideal depolarizer, and the relative weights are controlled by a single depolarization parameter as shown in Eq.~(\ref{eq:1par_a})\cite{Li_singleParam}. Under the assumptions of the TD model, the degrees of freedom in a MM are reduced from 16 to eight: one for $M_{00}$, three for the diattenuation and three for the retardance of dominant non-depolarizing MM $\widehat{\mathbf{m}}_0$, and one for the dominant coherency eigenvalue $\xi_0$ that controls the relative weights in the model.

The partial polarimetric method presented in this work relies on \emph{a priori} knowledge of the dominant process $\widehat{\mathbf{m}}_0$ and how it varies over the measured scattering geometries. Based on empirical observations, Fresnel reflection due to microfacets is used as the dominant process. The coordinate system transformations needed to calculate Fresnel reflection over the field of view of a measurement are described in Appendix~\ref{sect:fresnel}. 

For a TD MM, the contribution of $\mathbf{m}_{ID}$ outweighs the contribution of $\widehat{\mathbf{m}}_0$ when $\xi_0<0.625$. In this regime, subsurface scattering dominates the interaction, but the single largest coherent process can still be Fresnel reflection.

\subsection{Mueller Matrix Extrapolation}
A noise-free model for flux measurements $\mathbf{P}$ of a TD MM can be written as a linear system
\begin{equation}
    \mathbf{P}=\boldsymbol{\Phi}^{\mathrm{T}}\boldsymbol{\alpha}.
\end{equation}
Here $\boldsymbol{\Phi}$ is a matrix with rows that are the measurement matrix $\mathbf{W}$ applied to the dominant process $\widehat{\mathbf{m}}_0$ and ideal depolarizer $\mathbf{m}_{ID}$ from the TD model as in
\begin{equation}
    \boldsymbol{\Phi}=\begin{pmatrix}
    \mathbf{p}_0^{\mathrm{T}}\\
    \mathbf{p}_{ID}^{\mathrm{T}}
    \end{pmatrix}=\begin{pmatrix}
        [\mathbf{W}\widehat{\mathbf{m}}_0]^{\mathrm{T}}\\
        [\mathbf{W}\mathbf{m}_{ID}]^{\mathrm{T}}\\
    \end{pmatrix},  
    \label{eq:basis}
\end{equation}
and the elements of $\boldsymbol{\alpha}$ are the weights in the TD model as in
\begin{equation}
    \boldsymbol{\alpha}=\begin{pmatrix}
    \alpha_0\\ \alpha_{ID}
    \end{pmatrix}=\frac{4M_{00}}{3}\begin{pmatrix}
    \xi_0-\frac{1}{4}\\1-\xi_0
    \end{pmatrix}.
    \label{eq:alpha}
\end{equation}
It is worth reiterating that a benefit of the TD model is that the relative weights are controlled by a single depolarization parameter $\xi_0$, rather than varying independently. An estimate of the coefficients $\widetilde{\boldsymbol\alpha}$ can be recovered with the Moore-Penrose pseudoinverse of $\boldsymbol{\Phi}^\mathrm{T}$ as in
\begin{equation}
    \widetilde{\boldsymbol{\alpha}}=\left[\boldsymbol{\Phi}^{\mathrm{T}}\right]^+\mathbf{P}=\begin{pmatrix}
    
    \mathbf{W}\widehat{\mathbf{m}}_0 & \mathbf{W}\mathbf{m}_{ID}\end{pmatrix}^+\mathbf{P},
    \label{eq:pinv}
\end{equation}
where $\mathbf{P}$ here is a vector of noisy flux measurements.
Solving the system in Eq.~(\ref{eq:alpha}) for the model parameters, estimates for $\widetilde{\xi}_0$ and $\widetilde{M}_{00}$ are
\begin{eqnarray}
    \widetilde{\xi}_0=\frac{\frac{1}{4}+\widetilde{\alpha}_0/\widetilde{\alpha}_{ID}}{1+\widetilde{\alpha}_0/\widetilde{\alpha}_{ID}}, \, & \displaystyle \widetilde{M}_{00}=\frac{3\widetilde{\alpha}_0}{4\widetilde{\xi}_0-1}
    \label{eq:paramEst}
\end{eqnarray}
where $\widetilde{\alpha}_0$ and $\widetilde{\alpha}_{ID}$ are the elements of $\widetilde{\boldsymbol{\alpha}}$. $\widetilde{\xi}_0$ is the parameter of interest because it determines the fractional contributions of the dominant coherent process and the ideal depolarizer. This fractional contribution adjusts the depolarization of the MM which changes with scattering geometry, albedo, and surface texture\cite{kupinskiAoLP}. The estimated average reflectance $\widetilde{M}_{00}$ is ignored in the results because the polarimeters used were not calibrated to produce data in absolute radiometric units. Instead, the MM results will be normalized.

\subsection{Polarimeters}
\begin{figure}[H]
     \centering
     \begin{subfigure}[b]{0.43\textwidth}
         \centering
         \fbox{\includegraphics[trim= 10 900 10 1000, clip, height=0.7\textwidth]{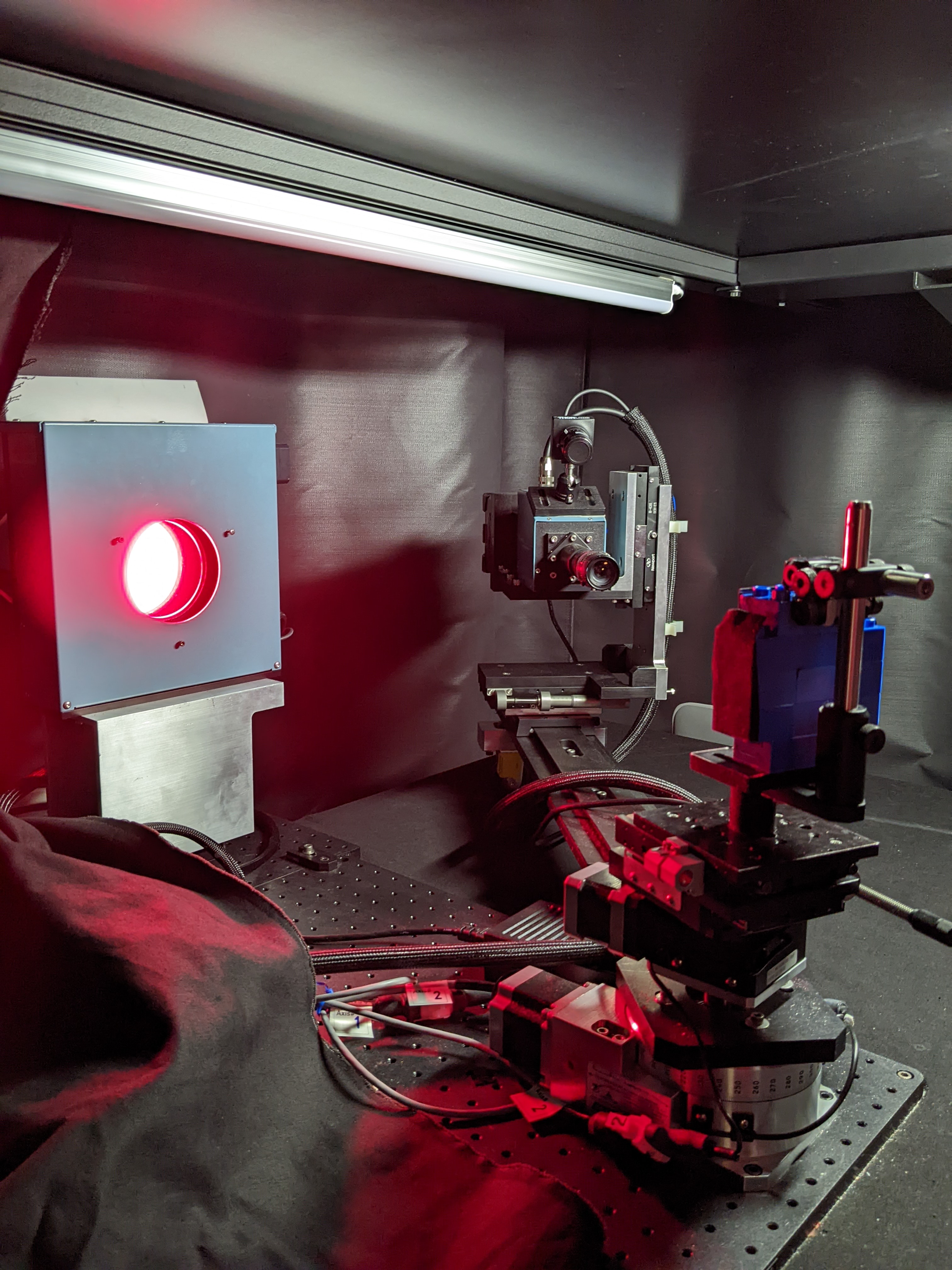}}
         \caption{Full Mueller polarimeter}
         \label{fig:rgb950}
     \end{subfigure}
     \begin{subfigure}[b]{0.43\textwidth}
         \centering
         \fbox{\includegraphics[trim= 10 700 10 1100, clip, height=0.7\textwidth]{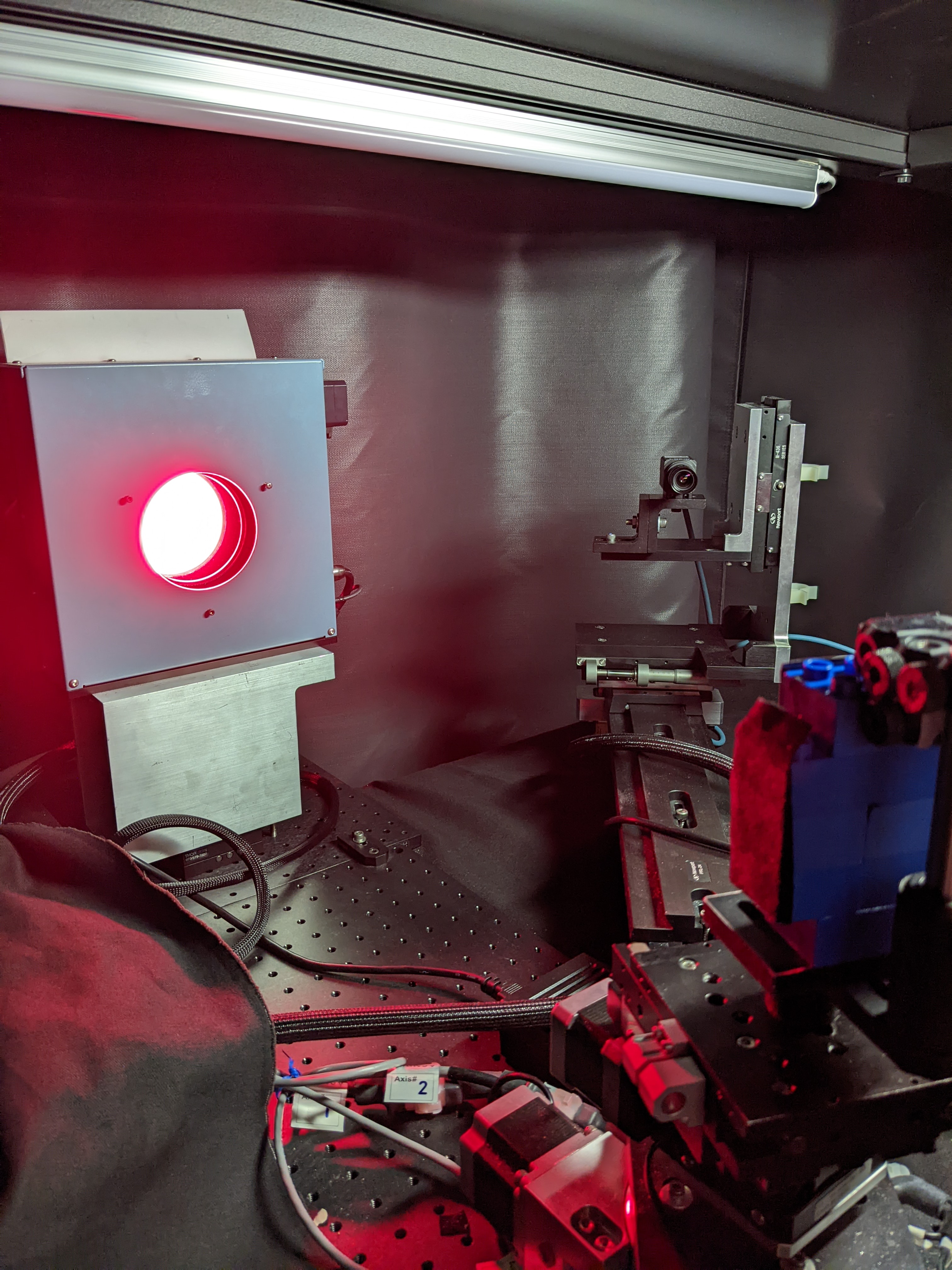}}
         \caption{Commercial linear Stokes camera}
         \label{fig:lucid}
     \end{subfigure}
        \caption{ (a) is the full Mueller imaging polarimeter referred to as the RGB950. It takes 40 polarimetric measurements. (b) is the goniometer and PSG of the RGB950 but with the PSA replaced by a commercial off-the-shelf linear Stokes camera. The Stokes camera takes 4 polarimetric measurements in a single snapshot with a static PSG state.  }
        \label{fig:polarimeters}
\end{figure}

The ground-truth MM images were taken using a full Mueller imaging polarimeter called the RGB950\cite{LopezTellez_rgb950}. The RGB950 (shown in Fig.~\ref{fig:rgb950}) is a dual rotating retarder polarimeter which reconstructs a MM image from a sequence of 40 polarimetric measurements at different retarder positions. Data was taken at two wavelengths: $662\pm 11.17$ nm (red) and $451\pm 9.83$ nm (blue). Thirty scattering geometries shown in Table \ref{tab:geom} were measured using a rotation stage for the sample and a goniometric swing arm for the camera. 

\begin{table}[H]
\caption{30 acquisition geometries specified on-axis where $\phi_i$ and $\phi_o$ are 0$^\circ$. For each angle between the sample surface normal and source, $\theta_i$, measurements are performed for six angles between the surface normal and the camera, $\theta_o$. The scattering geometries across the field of view of an image will have $\theta_i$, $\phi_i$, $\theta_o$, and $\phi_o$ that deviate from these on-axis values. Some acquisition geometries are omitted from analysis because exposure issues with the linear Stokes camera produced non-physical MM extrapolations.} 
\label{tab:geom}
\begin{center}       
\begin{tabular}{|l||l|l|l|l|l|l|}
\hline
\rule[-1ex]{0pt}{3.5ex}  $\theta_i$ & $\theta_{o,1}$ & $\theta_{o,2}$ & $\theta_{o,3}$ & $\theta_{o,4}$& $\theta_{o,5}$ & $\theta_{o,6}$  \\
\hline\hline
\rule[-1ex]{0pt}{3.5ex}  -10$^\circ$ & 10$^\circ$ & 20$^\circ$ & 30$^\circ$ & 40$^\circ$ & 50$^\circ$ & 60$^\circ$   \\
\hline
\rule[-1ex]{0pt}{3.5ex}  -25$^\circ$ & 15$^\circ$& 25$^\circ$& 35$^\circ$ & 45$^\circ$ & 55$^\circ$ & 65$^\circ$  \\
\hline
\rule[-1ex]{0pt}{3.5ex}  -40$^\circ$ & 20$^\circ$& 30$^\circ$& 40$^\circ$& 50$^\circ$& 60$^\circ$& 70$^\circ$  \\
\hline
\rule[-1ex]{0pt}{3.5ex} -55$^\circ$ & 25$^\circ$ & 35$^\circ$ &45$^\circ$ & 55$^\circ$& 65$^\circ$& 75$^\circ$  \\
\hline 
\rule[-1ex]{0pt}{3.5ex} -70$^\circ$ & 30$^\circ$ & 40$^\circ$& 50$^\circ$ & 60$^\circ$ & 70$^\circ$ & 80$^\circ$ \\
\hline 
\end{tabular}
\end{center}
\end{table}

The linear partial polarimetric experiment was performed using a Sony Triton 5.0MP Polarization Camera shown in Fig.~\ref{fig:lucid}. This camera has an array of micropolarizers in front of the detector elements so 4 polarimetric measurements (three of which are linearly-independent) are taken simultaneously at the cost of spatial resolution. The rank three measurement matrix of this system is underdetermined for full Mueller polarimetry, but overdetermined for recovering the two unknown degrees of freedom in the TD model. Designing polarimeters to be overdetermined systems is a well-established practice in the literature\cite{Smith:02,Twietmeyer:08,Vaughn:08}. The polarization state generator of the RGB950 was used to keep the illumination consistent, but only horizontal linear polarization was used with the Stokes camera measurements. In principle, unpolarized illumination would also work. In this work, the calibration and operation software provided by the manufacturer are used. This could lead to exaggerated errors in the extrapolation results.

\subsection{Samples}

   \begin{figure}[H]
         \centering
         \includegraphics[width=.4\textwidth]{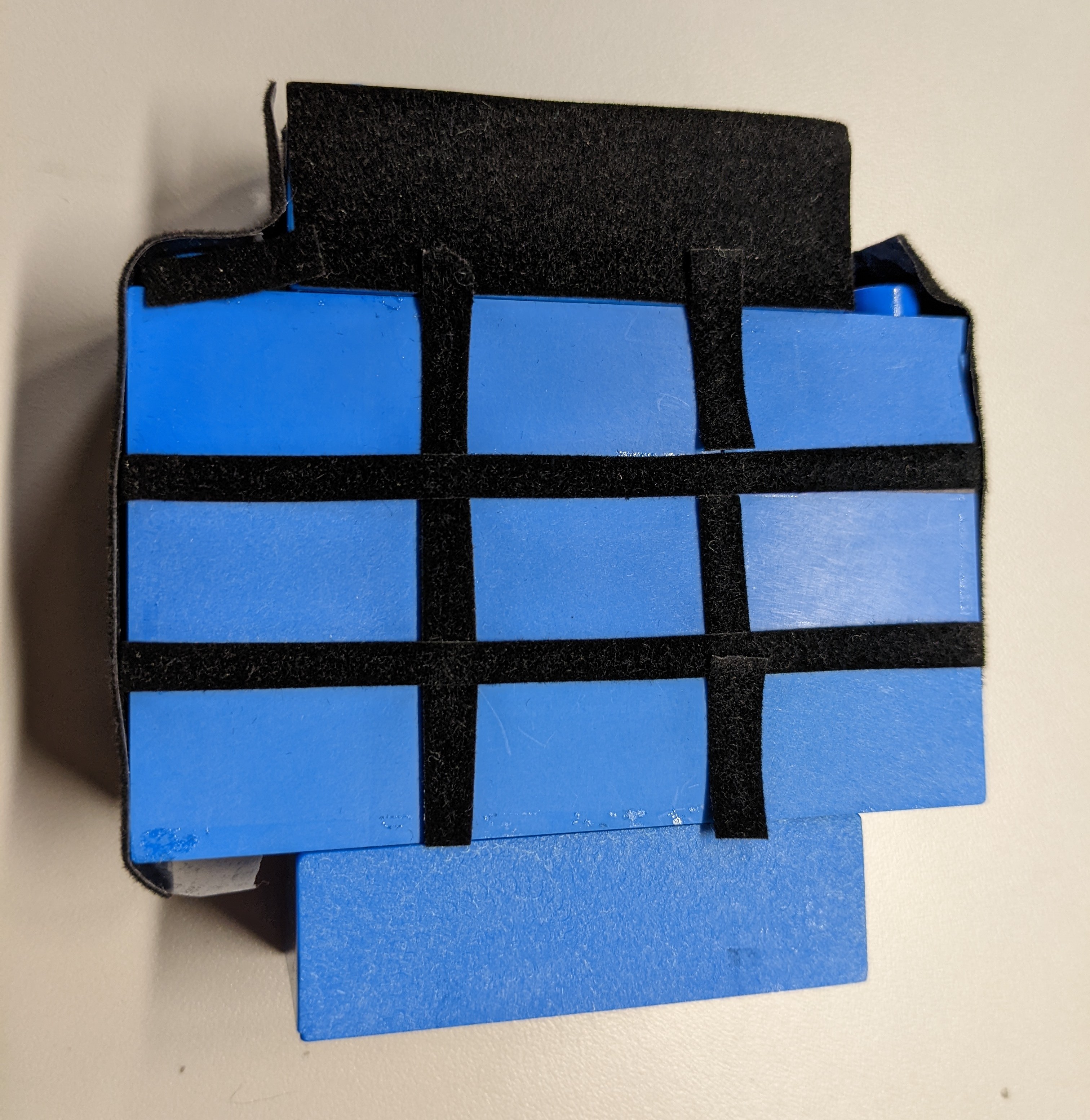}
        \caption{A tower of blue plastic LEGO bricks that have each been sanded with different grits of sandpaper. This represents a group of objects with similar properties and albedos for a given wavelength but with different surface textures. The roughness averages (Ra) in microns for each brick are top: 0.49, 0.56, 3.35, middle: 3.55, 2.62, 0.35, bottom: 1.68, 1.26, 6.32.}
        \label{fig:legos}
\end{figure}

The samples measured in this work are a collection of blue LEGO bricks shown in Fig.~\ref{fig:legos}. These are the same LEGOs used by Li and Kupinski and are a collection of individual objects with the same material properties and albedo but with varying texture\cite{Li_singleParam}. The surface roughness of each brick was measured using a white light interferometer. The roughness averages (Ra) in microns for each brick are top: 0.49, 0.56, 3.35, middle: 3.55, 2.62, 0.35, bottom: 1.68, 1.26, 6.32. Since the bricks are blue, the different wavelengths represent different albedo cases: 662 nm illumination is low albedo and 451 nm illumination is high albedo. Umov's effect states that the amount of depolarization is expected to trend positively with albedo, so these albedo cases also represent cases with different amounts of depolarization\cite{umow1905chromatische}.

\section{Results}


\begin{figure}[!th]
\centering
    \makebox[5pt]{{\includegraphics[width=0.6\columnwidth]{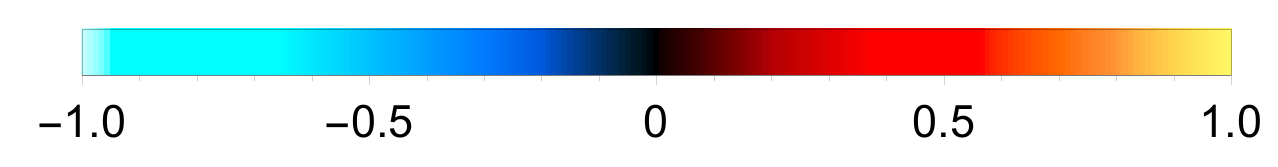}}}\\
              \centering
    \begin{subfigure}[b]{0.49\textwidth}
         \centering
         \includegraphics[trim= 0 0 60 0, clip,width=\textwidth]{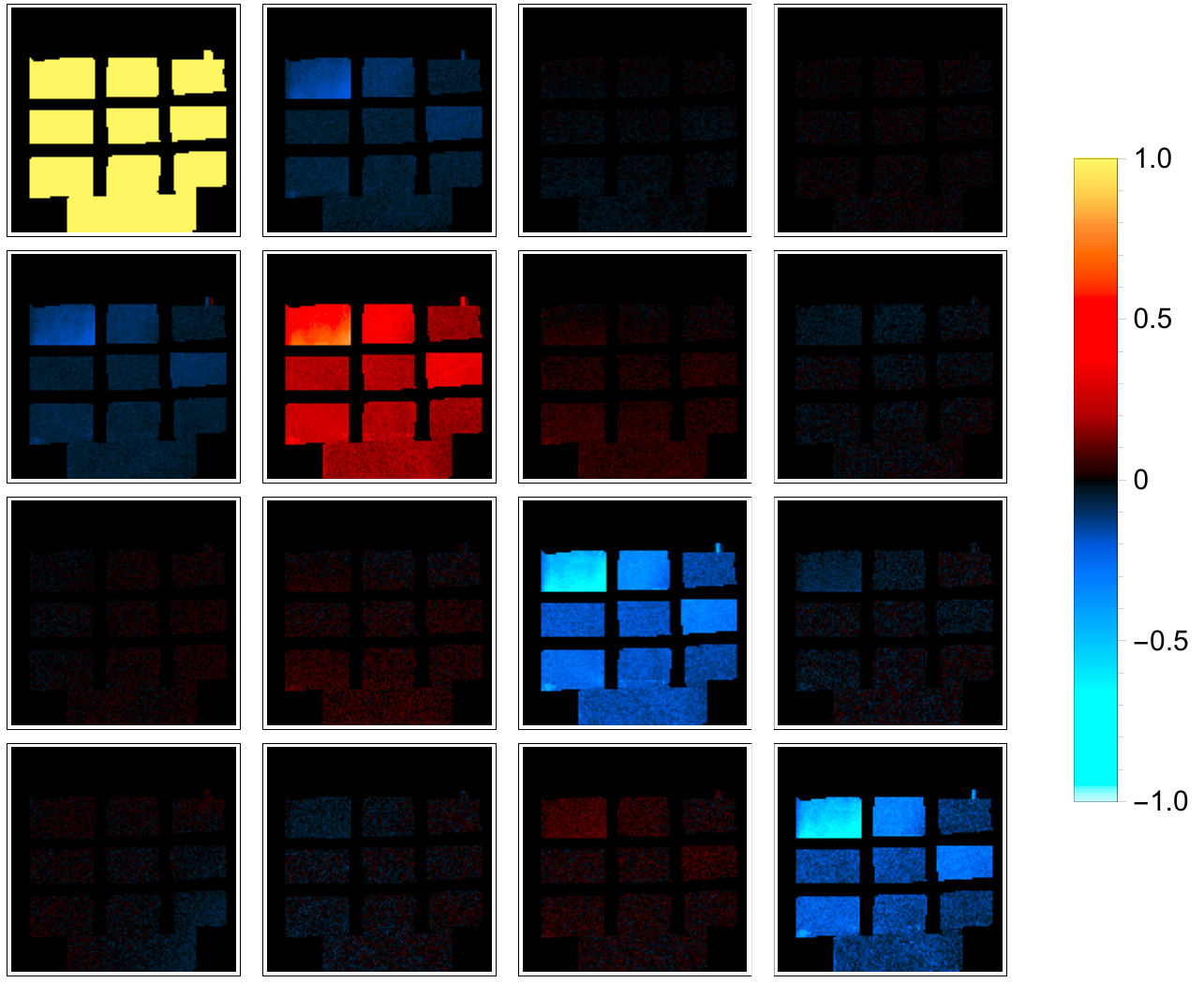}
         \caption{Full reconstruction 451 nm}
         \label{fig:true451}
    \end{subfigure}
         \hfill     
    \begin{subfigure}[b]{0.49\textwidth}
         \centering
         \includegraphics[trim= 0 0 60 0, clip,width=\textwidth]{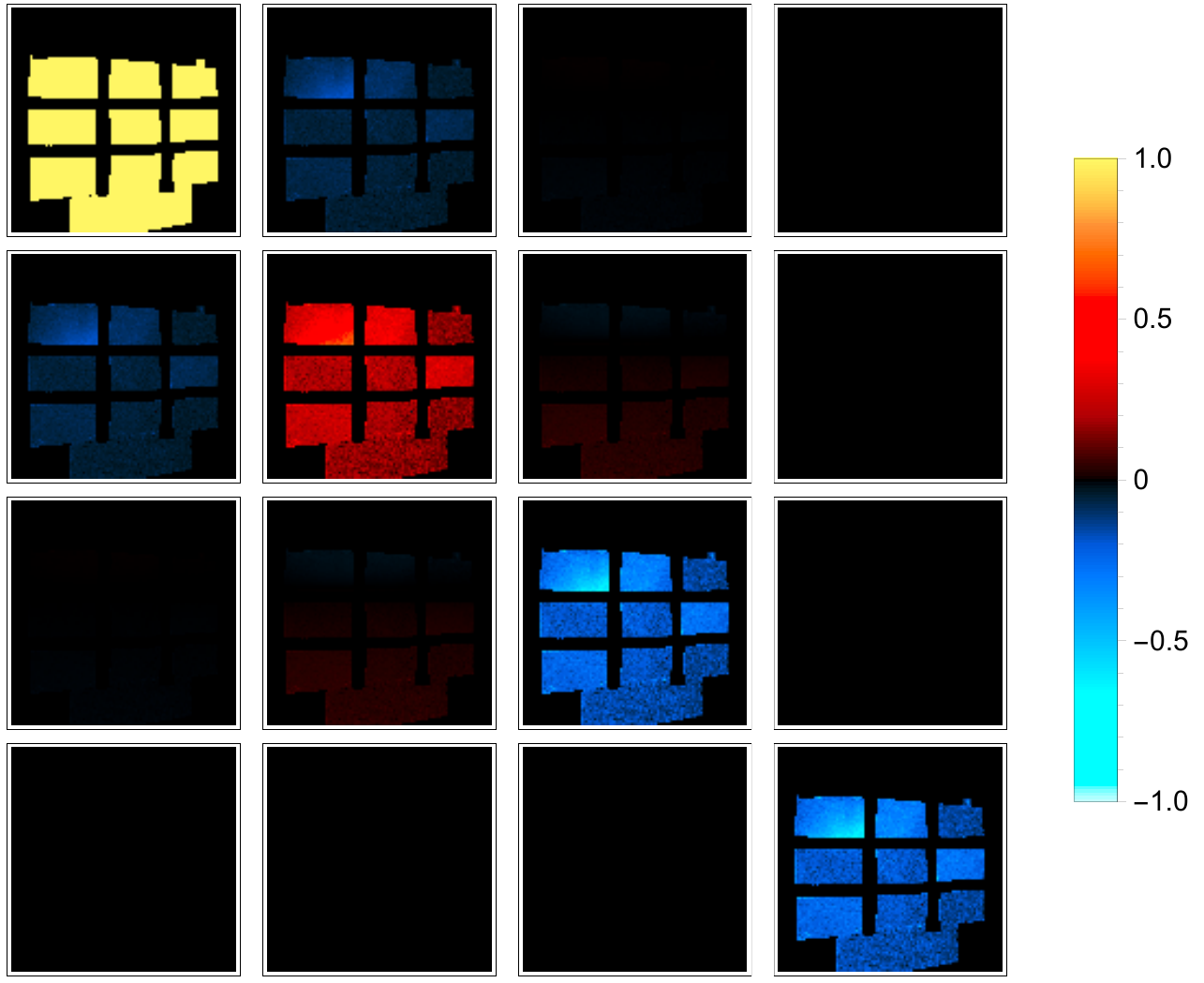}
         \caption{Extrapolation 451 nm}
         \label{fig:extrap451}
    \end{subfigure}
     \centering
    \begin{subfigure}[b]{0.49\textwidth}
         \centering
         \includegraphics[trim= 0 0 60 0, clip,width=\textwidth]{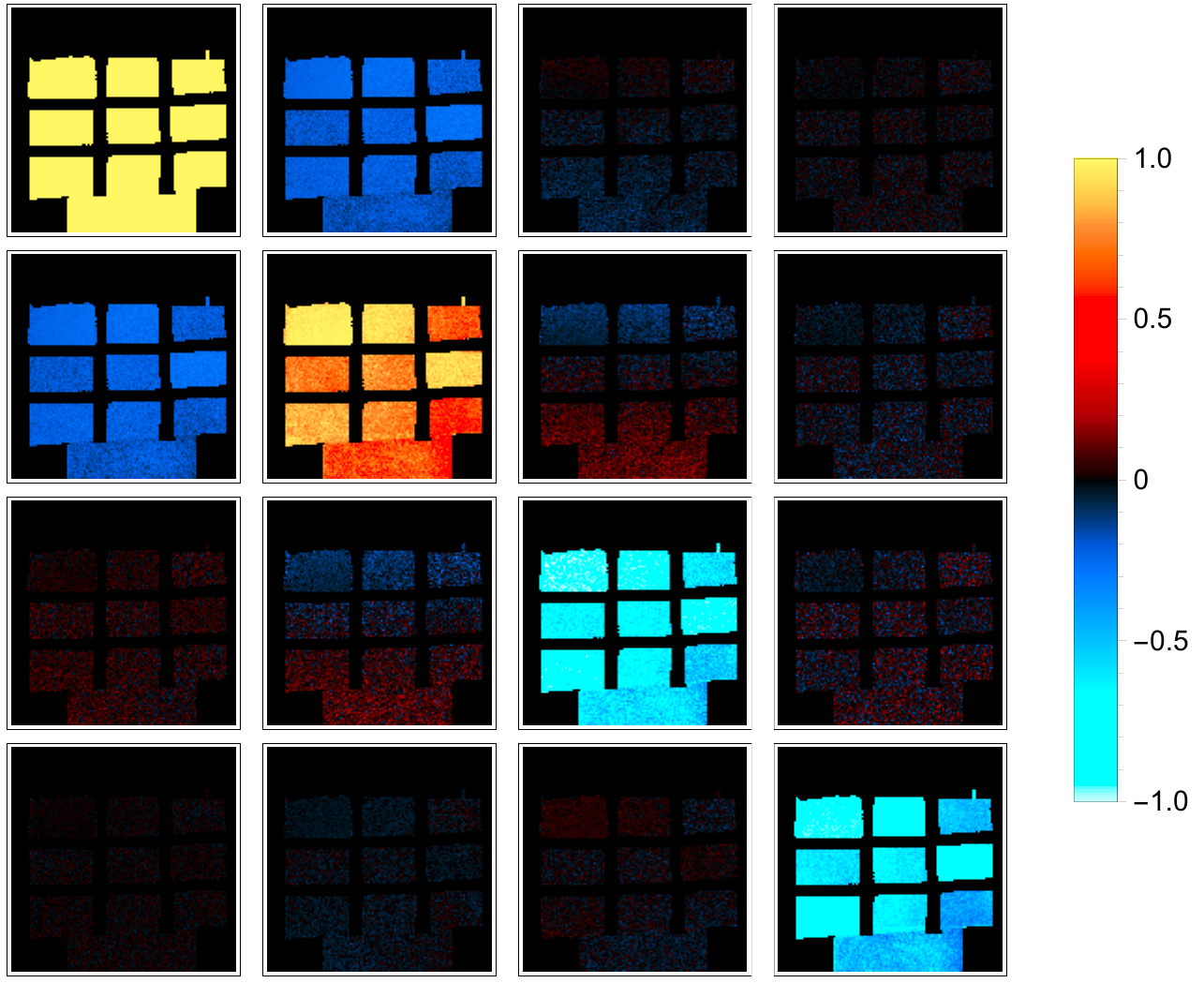}
         \caption{Full reconstruction 662 nm}
         \label{fig:true662}
    \end{subfigure}
         \hfill     
    \begin{subfigure}[b]{0.49\textwidth}
         \centering
         \includegraphics[trim= 0 0 60 0, clip,width=\textwidth]{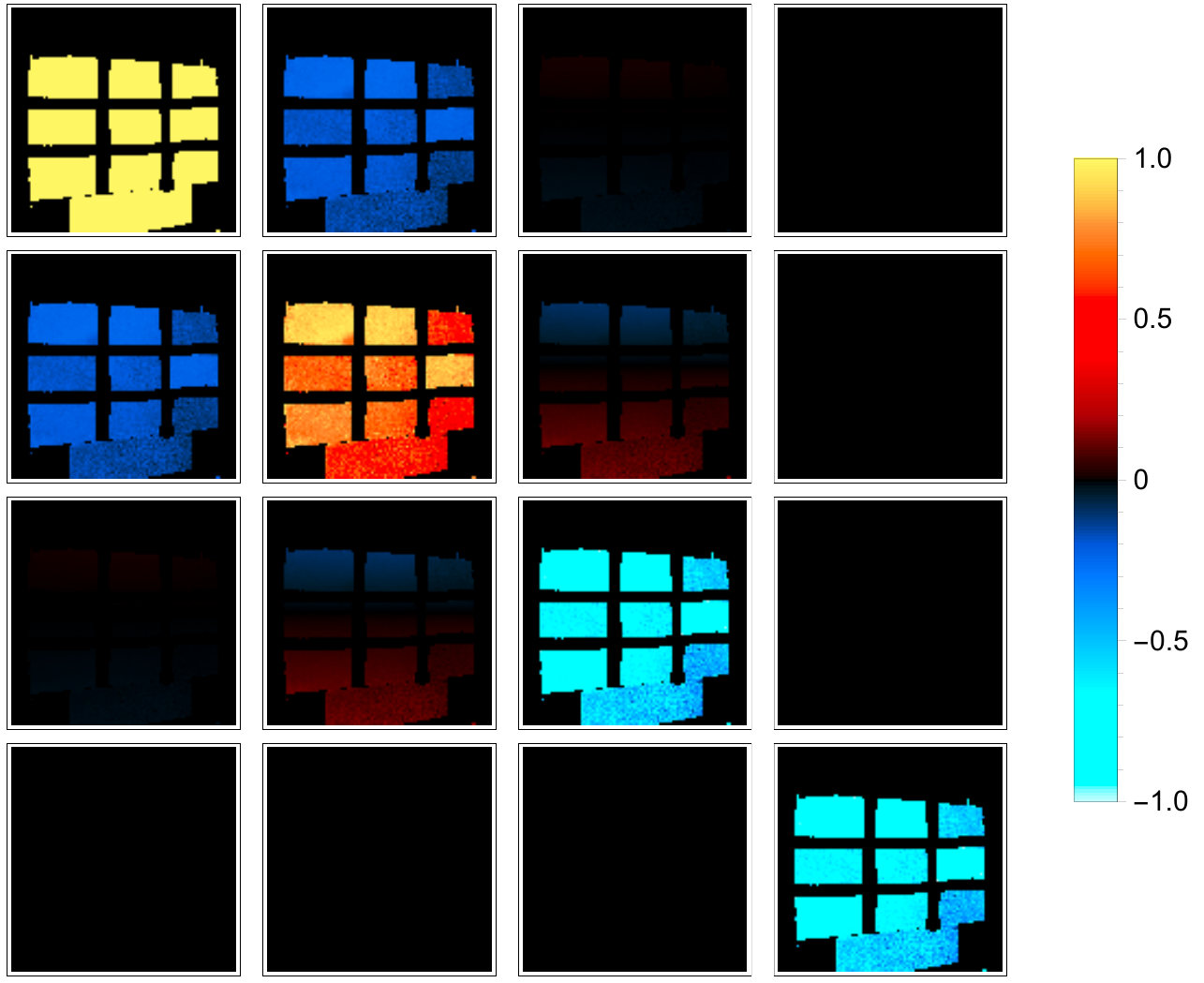}
         \caption{Extrapolation 662 nm}
         \label{fig:extrap662}
    \end{subfigure}
    \caption{ Comparison of the MM image results at 451 nm of (a) full reconstruction with 40 polarimetric measurements to (b) MM image results of partial polarimetric extrapolation (Video 1) and at 662 nm of (c) full reconstruction and (d) extrapolation (Video 2). The still images are for the geometry $\theta_i=-25^{\circ}$, $\theta_o=25^{\circ}$. (Video 1, MP4, 3MB; Video 2, MP4, 4MB). }\label{vid:MMs}
\end{figure}

Video~\ref{vid:MMs} shows a comparison of MM images calculated using traditional Mueller polarimetry with 40 polarimetric measurements versus partial polarimetric extrapolations using 4 polarimetric measurements. Depolarization can be qualitatively observed by comparing the magnitude of $M_{00}$ to other matrix elements: regions of the image where all matrix elements have a smaller magnitude than $M_{00}$ have larger depolarization. Depolarization is expected to be stronger for the high albedo case of blue bricks under blue illumination, and this can be seen by the relatively lower magnitudes across the field of view in both the reconstruction and extrapolation. Likewise, the expectation of lower depolarization is met for the low albedo case of the blue bricks under red illumination. The trend of increased depolarization with surface roughness is also captured by the extrapolation.

\subsection{Error in Dominant Eigenvalue Estimate}

\begin{figure}[H]
\centering
    \makebox[5pt]{{\includegraphics[width=0.6\columnwidth]{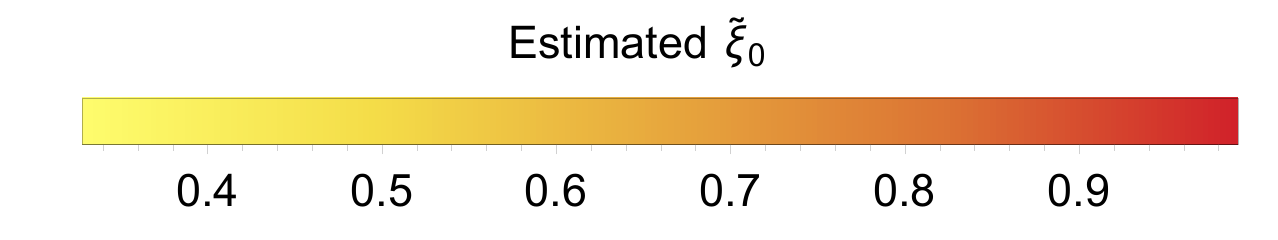}}}\\
    \centering
    \begin{subfigure}[b]{0.49\textwidth}
         \centering
         \includegraphics[width=\textwidth]{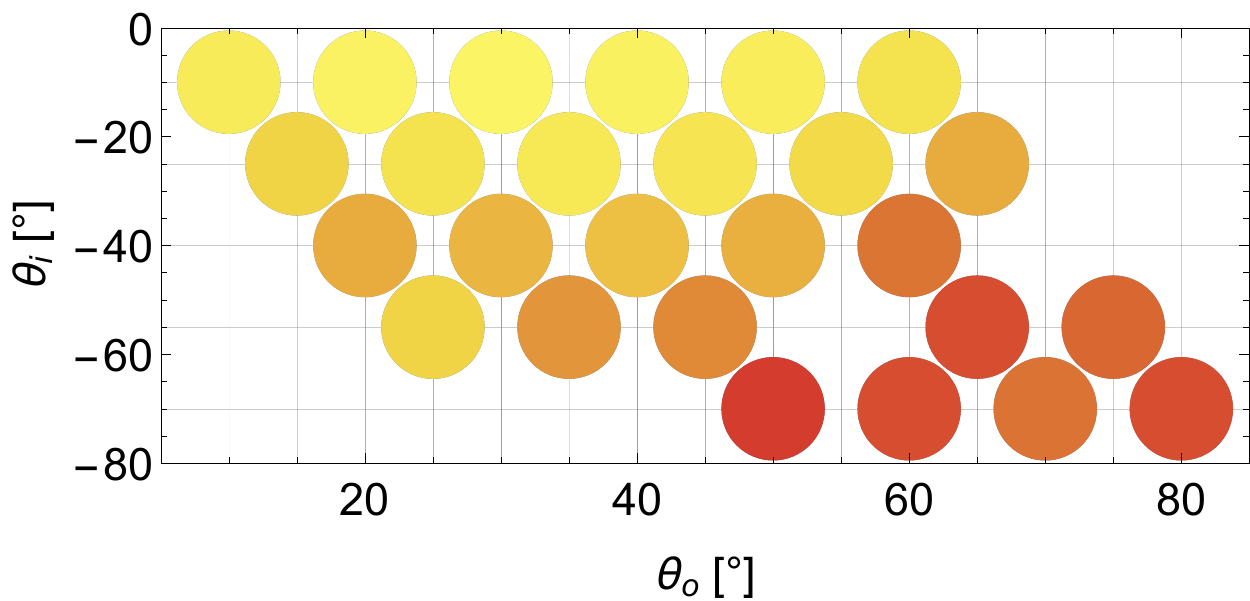}
         \caption{Smoothest brick, 451 nm}
         \label{fig:smoothXi0est_451}
    \end{subfigure}
         \hfill     
    \begin{subfigure}[b]{0.49\textwidth}
         \centering
         \includegraphics[width=\textwidth]{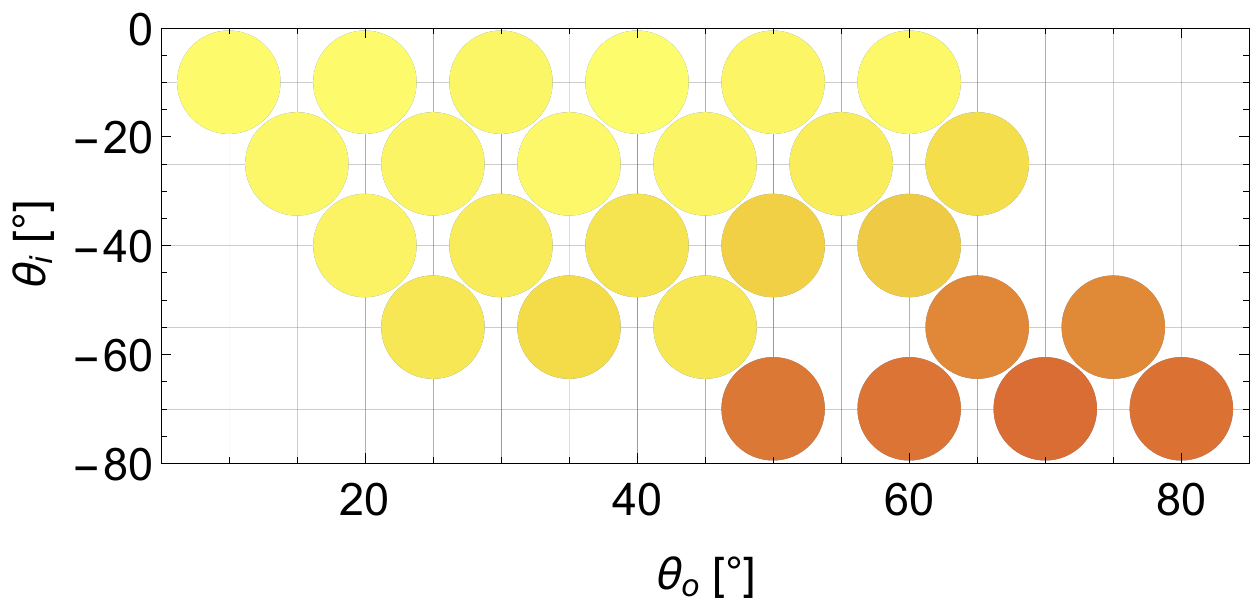}
         \caption{Roughest brick, 451 nm}
         \label{fig:roughXi0est_451}
    \end{subfigure}
    \centering
    \begin{subfigure}[b]{0.49\textwidth}
         \centering
         \includegraphics[width=\textwidth]{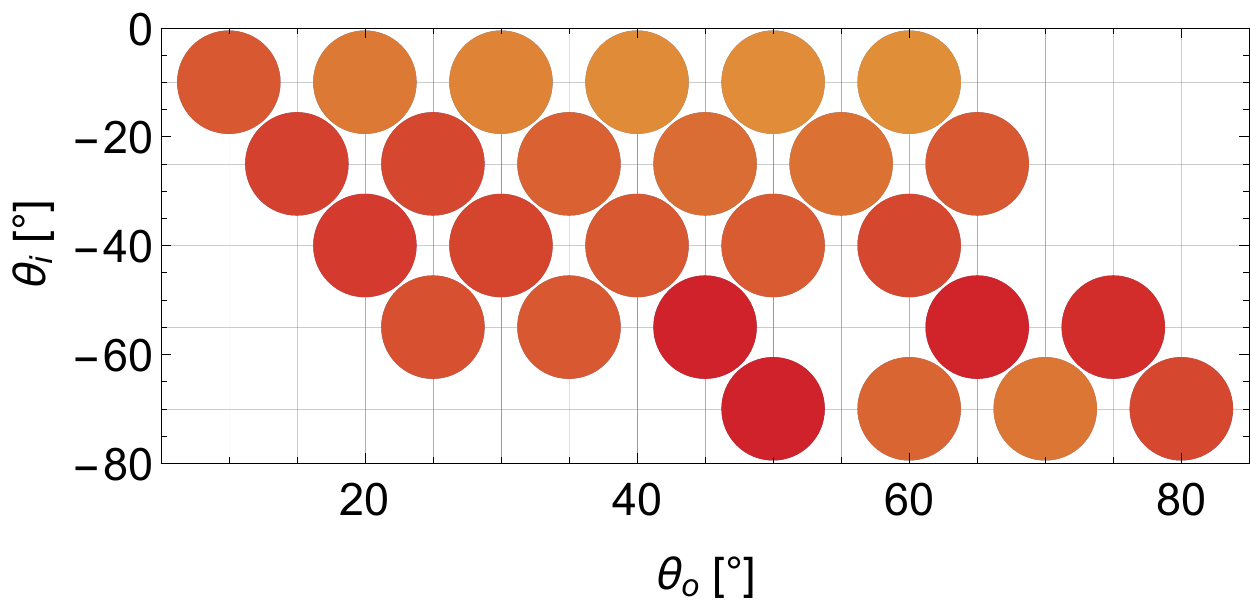}
         \caption{Smoothest brick, 662 nm}
         \label{fig:smoothXi0est_662}
    \end{subfigure}
         \hfill     
    \begin{subfigure}[b]{0.49\textwidth}
         \centering
         \includegraphics[width=\textwidth]{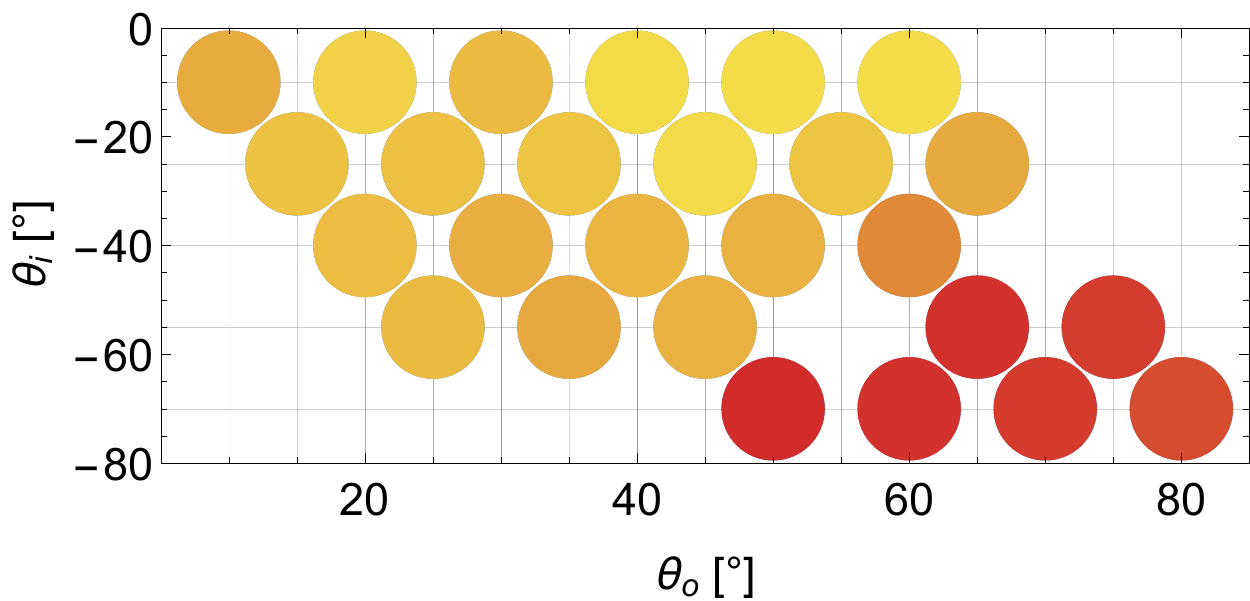}
         \caption{Roughest brick, 662 nm}
         \label{fig:roughXi0est_662}
    \end{subfigure}
    \caption{The estimate $\widetilde{\xi}_0$ calculated as in Eq.~(\ref{eq:paramEst}) for two different brick textures at both wavelengths. Geometries at which the dynamic range of the linear Stokes camera caused non-physical MM extrapolations are omitted. }\label{fig:xi0est}
\end{figure}

Figure \ref{fig:xi0est} shows the estimated values of $\widetilde{\xi}_0$ for the smoothest and roughest textured bricks. Larger values correspond to a larger estimated contribution of the dominant process, or equivalently lower depolarization. For both brick textures, estimates of $\widetilde{\xi}_0$ are larger in the low albedo case of 662 nm illumination than in the high albedo case of 451 nm illumination. This is in agreement with expectations from Umov's effect where depolarization trends positively with albedo. Furthermore, for both wavelengths, estimates of $\widetilde{\xi}_0$ are larger for the smooth brick and smaller for the rough brick. This trend matches the expectation of a rougher texture resulting in higher depolarization.

\begin{figure}[H]
\centering
    \makebox[5pt]{{\includegraphics[width=0.6\columnwidth]{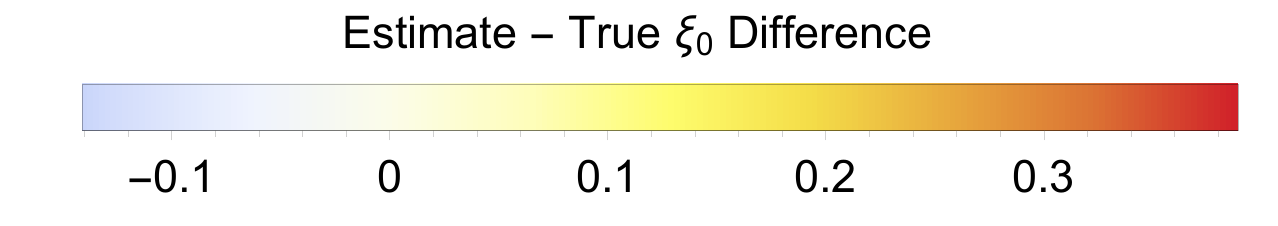}}}\\
    \centering
    \begin{subfigure}[b]{0.49\textwidth}
         \centering
         \includegraphics[width=\textwidth]{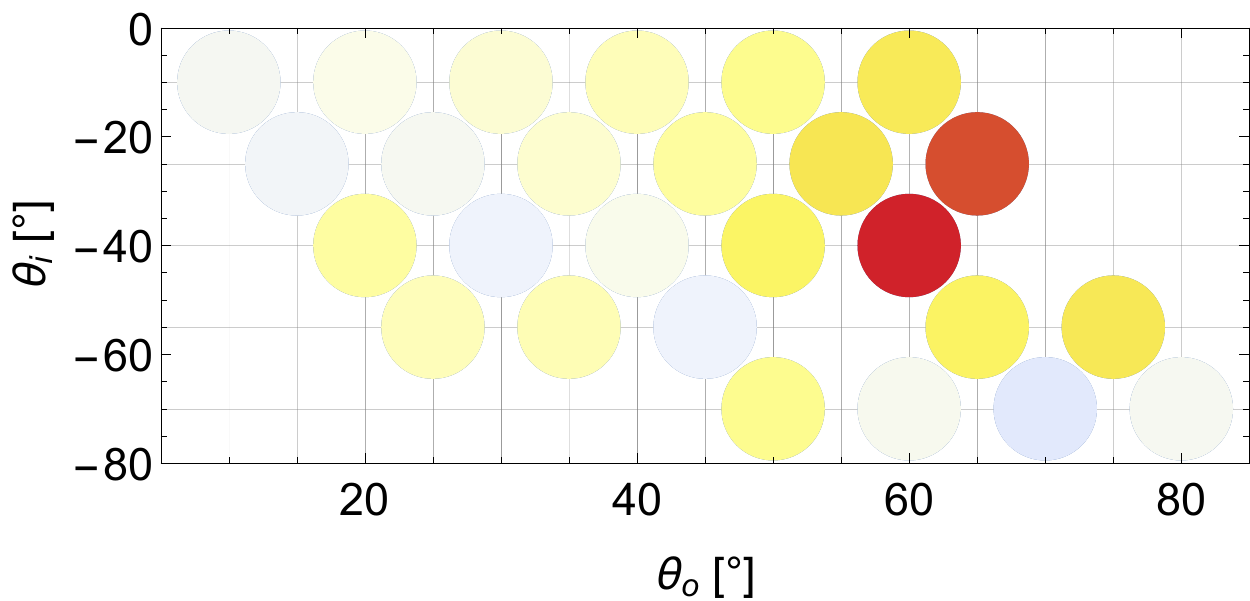}
         \caption{Smoothest brick, 451 nm}
         \label{fig:smoothXi0_451}
    \end{subfigure}
         \hfill     
    \begin{subfigure}[b]{0.49\textwidth}
         \centering
         \includegraphics[width=\textwidth]{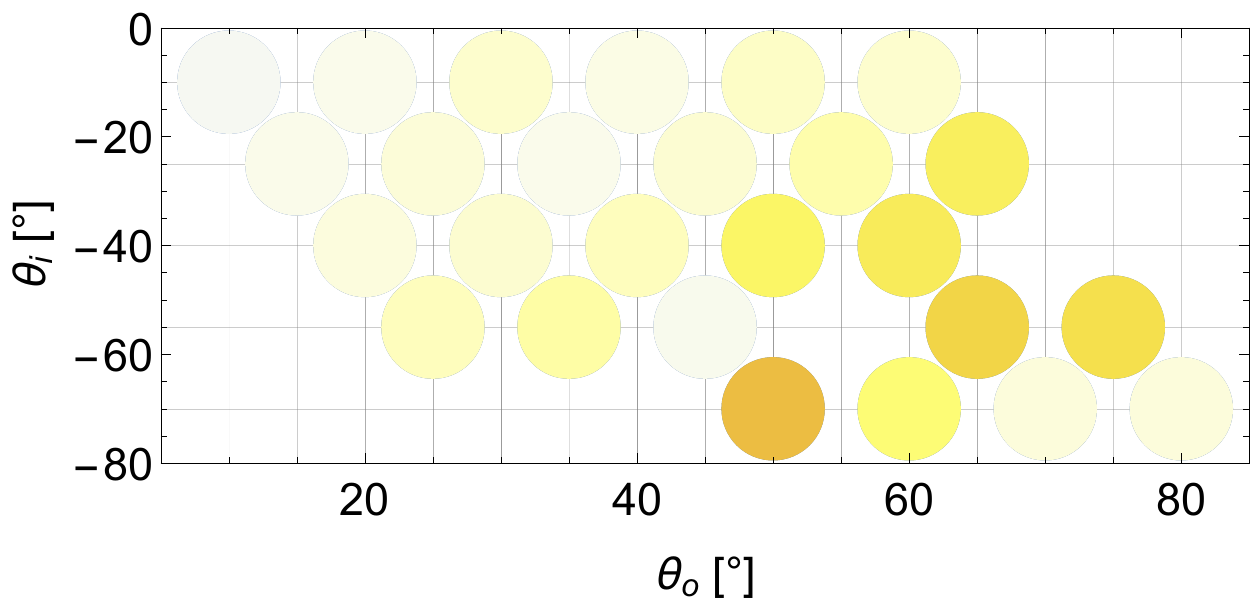}
         \caption{Roughest brick, 451 nm}
         \label{fig:roughXi0_451}
    \end{subfigure}
    \centering
    \begin{subfigure}[b]{0.49\textwidth}
         \centering
         \includegraphics[width=\textwidth]{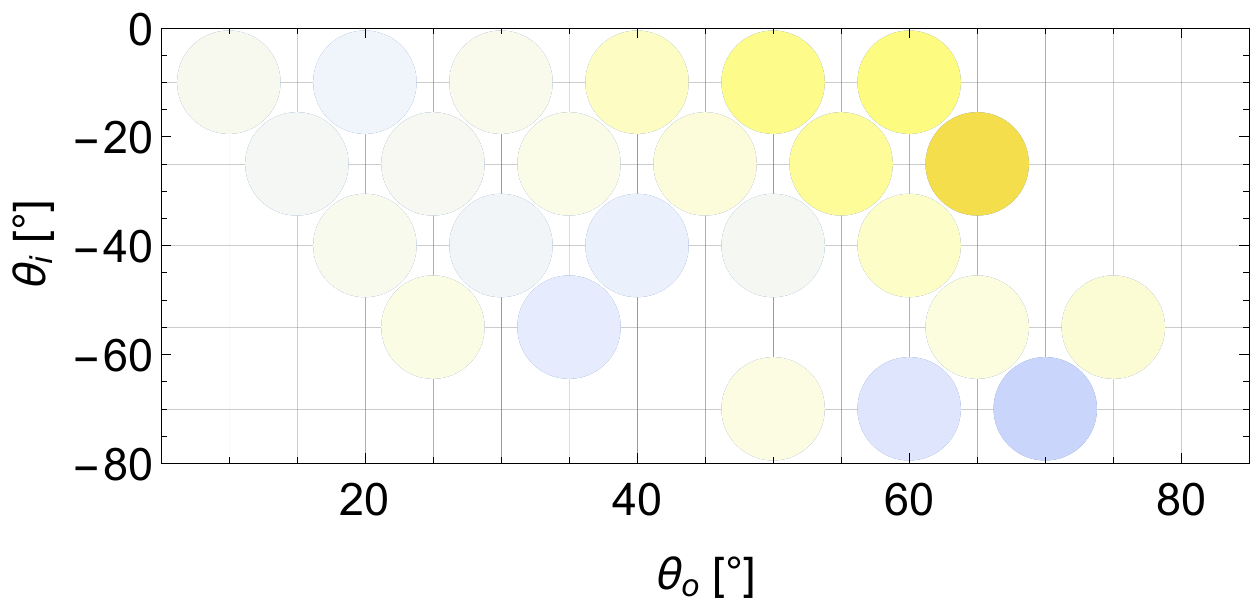}
         \caption{Smoothest brick, 662 nm}
         \label{fig:smoothXi0_662}
    \end{subfigure}
         \hfill     
    \begin{subfigure}[b]{0.49\textwidth}
         \centering
         \includegraphics[width=\textwidth]{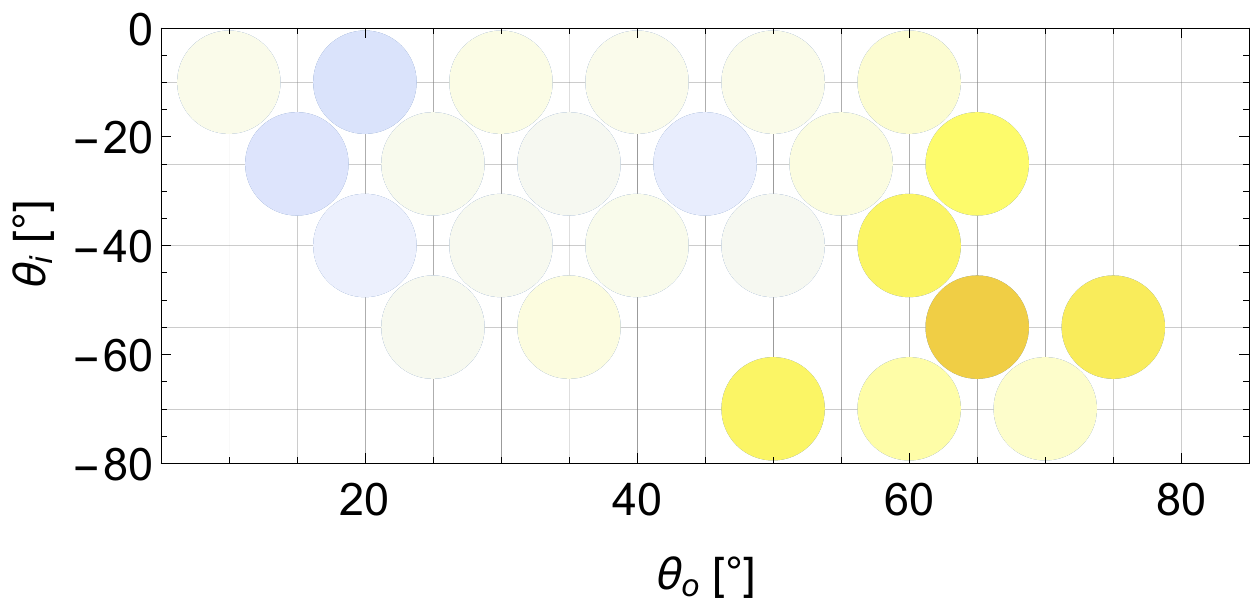}
         \caption{Roughest brick, 662 nm}
         \label{fig:roughXi0_662}
    \end{subfigure}
    \caption{The estimate $\widetilde{\xi}_0$ calculated as in Eq.~(\ref{eq:paramEst}) minus the true $\xi_0$ versus acquisition geometry. When the difference is positive valued, the contribution of the dominant non-depolarizing term is overestimated. Geometries at which the dynamic range of the linear Stokes camera caused non-physical MM extrapolations are omitted. }\label{fig:xi0diff}
\end{figure}

Figure~\ref{fig:xi0diff} shows the difference between the true value of $\xi_0$ from the complete MM reconstruction and the estimated $\widetilde{\xi}_0$ from the linear Stokes measurements. Positive-valued differences correspond to an overestimation of the dominant non-depolarizing process or equivalently an underestimation of the amount of depolarization.

Extrapolations at 451 nm tended to overestimate $\widetilde{\xi}_0$ at more geometries than at 662 nm. 662 nm is the low-albedo case, where Umov's effect indicates that depolarization is lower, so it is possible that the method is most successful for low-depolarization cases.

\subsection{Simulated Flux Vectors}

\begin{figure}[H]
    \centering
    \begin{subfigure}[b]{0.49\textwidth}
         \centering
         \includegraphics[width=\textwidth]{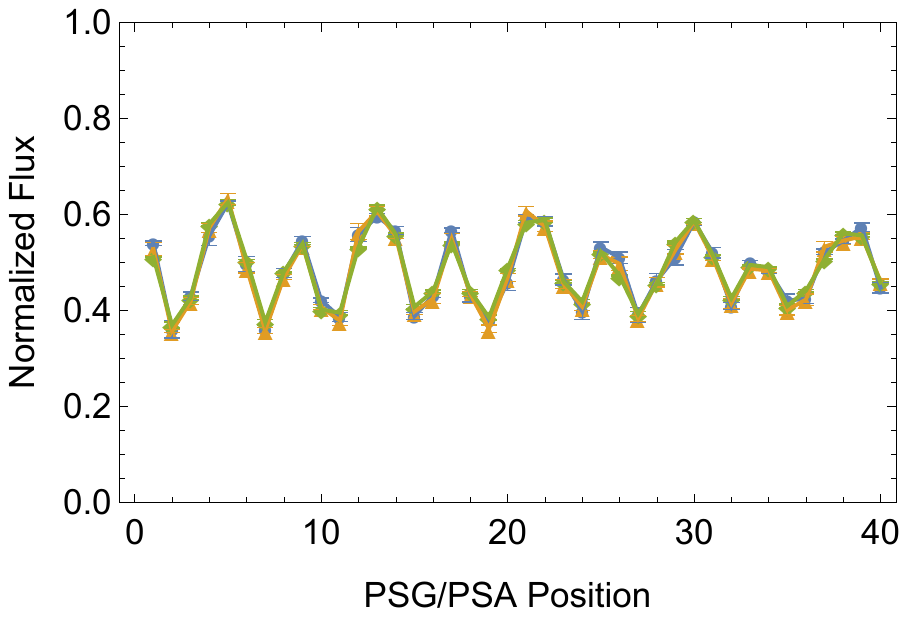}
         \caption{Smoothest brick, 451 nm}
         \label{fig:smoothFlux_451}
    \end{subfigure}
         \hfill     
    \begin{subfigure}[b]{0.49\textwidth}
         \centering
         \includegraphics[width=\textwidth]{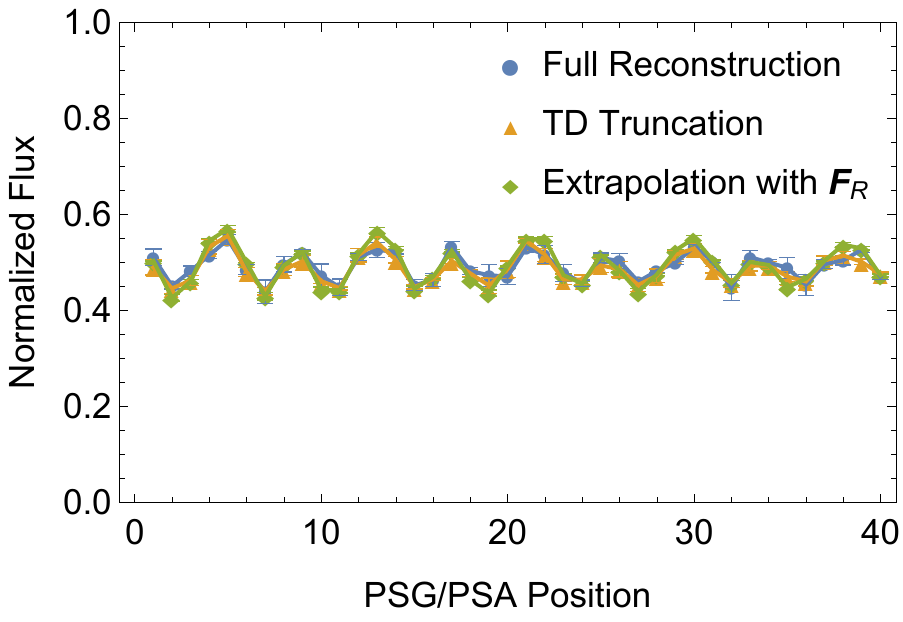}
         \caption{Roughest brick, 451 nm}
         \label{fig:roughFlux_451}
    \end{subfigure}
    \centering
    \begin{subfigure}[b]{0.49\textwidth}
         \centering
         \includegraphics[width=\textwidth]{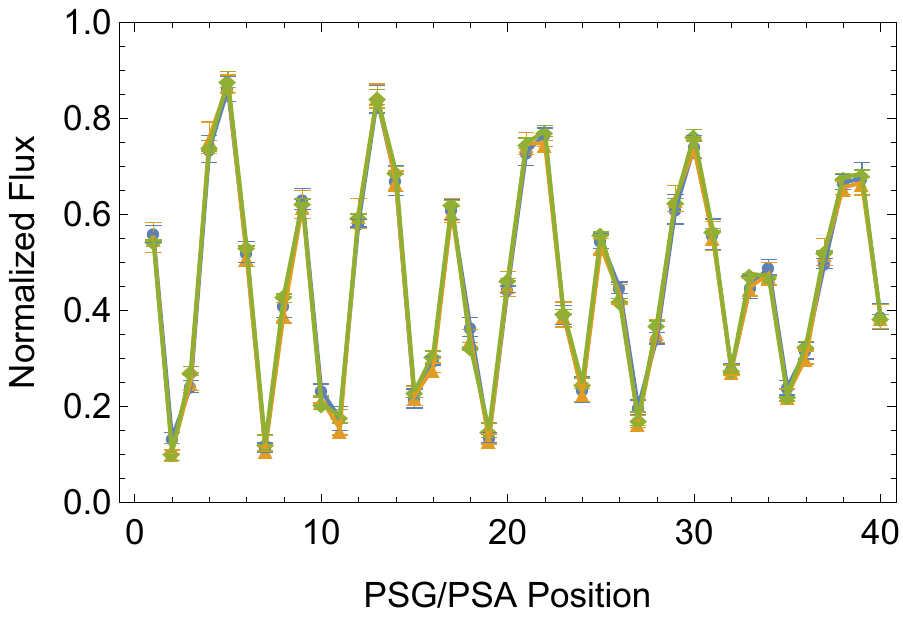}
         \caption{Smoothest brick, 662 nm}
         \label{fig:smoothFlux_662}
    \end{subfigure}
         \hfill     
    \begin{subfigure}[b]{0.49\textwidth}
         \centering
         \includegraphics[width=\textwidth]{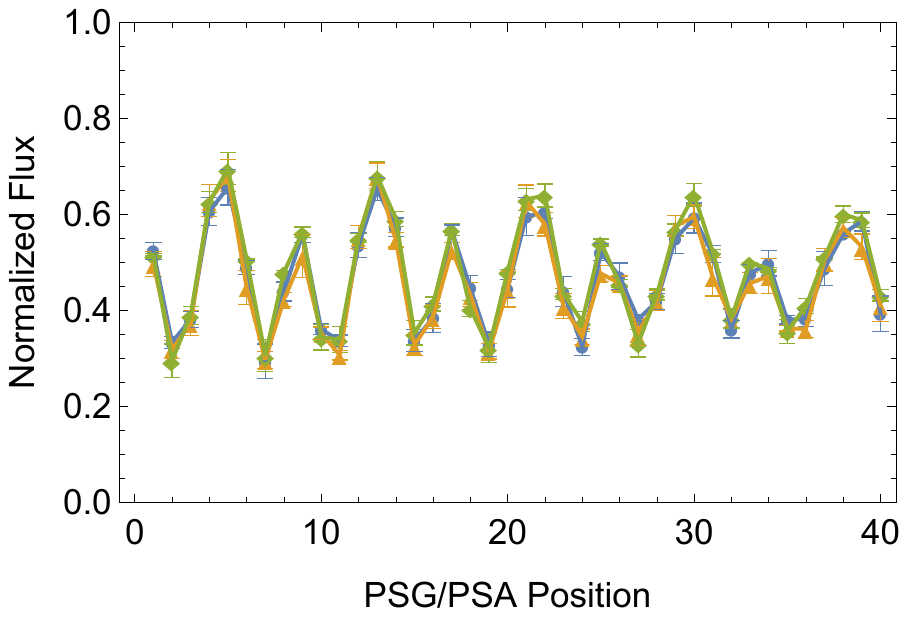}
         \caption{Roughest brick, 662 nm}
         \label{fig:roughFlux_662}
    \end{subfigure}
    \caption{ Flux vectors at 451 nm of (a) the smoothest brick and (b) the roughest brick (Video 3) and at 662 nm of (c) smoothest brick and (d) the roughest brick (Video 4). Flux vectors are calculated by averaging the normalized MM on a $2\times2$ pixel ROI then applying $\mathbf{W}_{40}$ to simulate what the RGB950 would measure. The error bars are $\pm$1 standard deviation in the ROI. The flux vectors shown here are for the full MM reconstructions (blue), the nearest TD approximation of the reconstruction (orange), and the extrapolated MMs (green) at $\theta_i=-25^{\circ}$, $\theta_o=25^{\circ}$. Geometries at which the dynamic range of the linear Stokes camera caused non-physical MM extrapolations are omitted from the videos. (Video 3, MP4, 9MB; Video 4, MP4, 9MB). }\label{vid:fluxVectors}
\end{figure}

To compare the MMs, the measurement matrix of the RGB950 $\mathbf{W}_{40}$ is applied to a $2\times2$ pixel average of the normalized extrapolated MM and the full reconstructed MM to simulate the flux measurements that the RGB950 would take. Additionally, the measurement noise is indicated by the standard deviation error bar on each flux measurement. The resulting flux vectors are shown in Vid.~(\ref{vid:fluxVectors}). Also shown in Vid.~(\ref{vid:fluxVectors}) is the nearest TD approximation of the reconstructed MM. This is calculated by setting the three smallest coherency eigenvalues to $\frac{1-\xi_0}{3}$. This TD approximation has the exact correct dominant process $\widehat{\mathbf{m}}_0$ and represents the best possible extrapolation based on a TD model. Flux vectors where the TD approximation and extrapolation show a similar deviation from the full reconstruction could indicate that the TD assumption is not valid. However, the movie of flux vectors over acquisition geometry does not show consistent disagreement between the reconstruction and the TD approximation, where the extrapolation also deviates. The measurements at 662 nm (the low albedo case) exhibit larger error bars for both the smooth and rough brick as compared to measurements at 451 nm (high albedo). This matches expectation since, despite larger polarization modulation for low albedo per Umov's effect, the overall amount of light is lower\cite{umow1905chromatische}. The largest realizations of measurement noise occur for the rough brick at scattering geometries near those with non-physical results but are not yet themselves non-physical. However, for all other measurements the disagreement between the extrapolations and the full reconstruction is larger than the error bars. This means that errors in the extrapolation are more likely the result of discrepancies in the assumed dominant process.

\begin{figure}[H]
    \centering
    \begin{subfigure}[b]{0.49\textwidth}
         \centering
         \includegraphics[width=\textwidth]{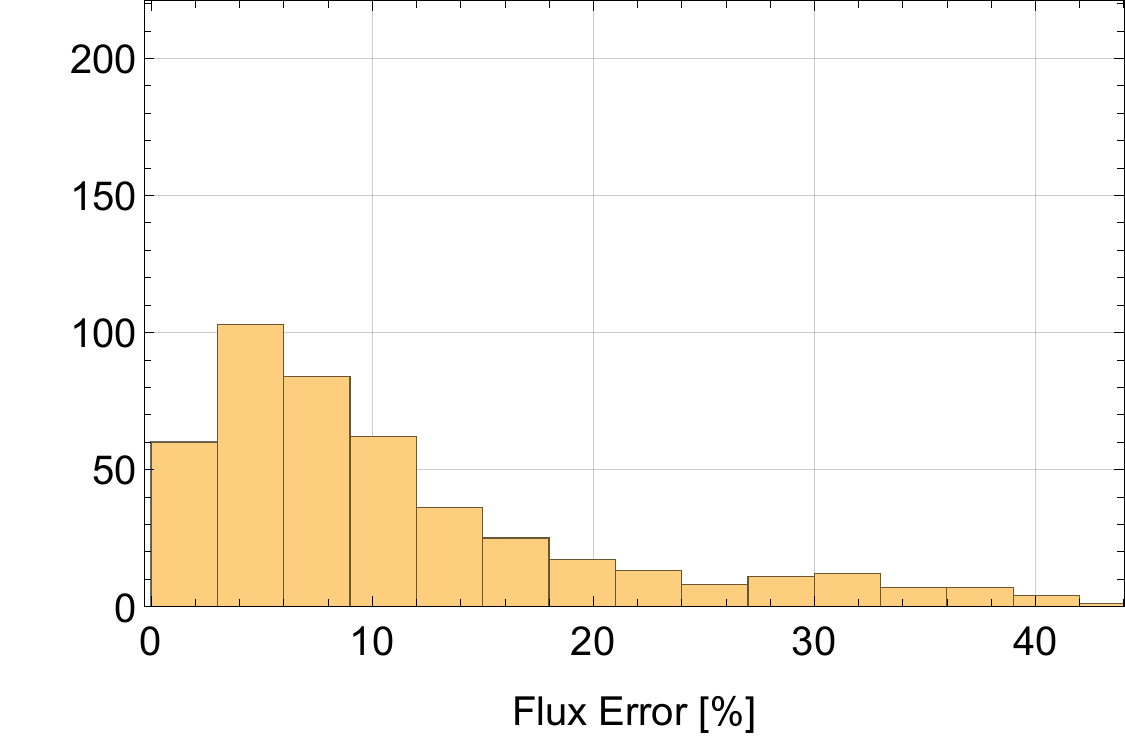}
         \caption{Extrapolation using $\mathbf{F}^R$}
         \label{fig:histogramExtrap}
    \end{subfigure}
         \hfill     
    \begin{subfigure}[b]{0.49\textwidth}
         \centering
         \includegraphics[width=\textwidth]{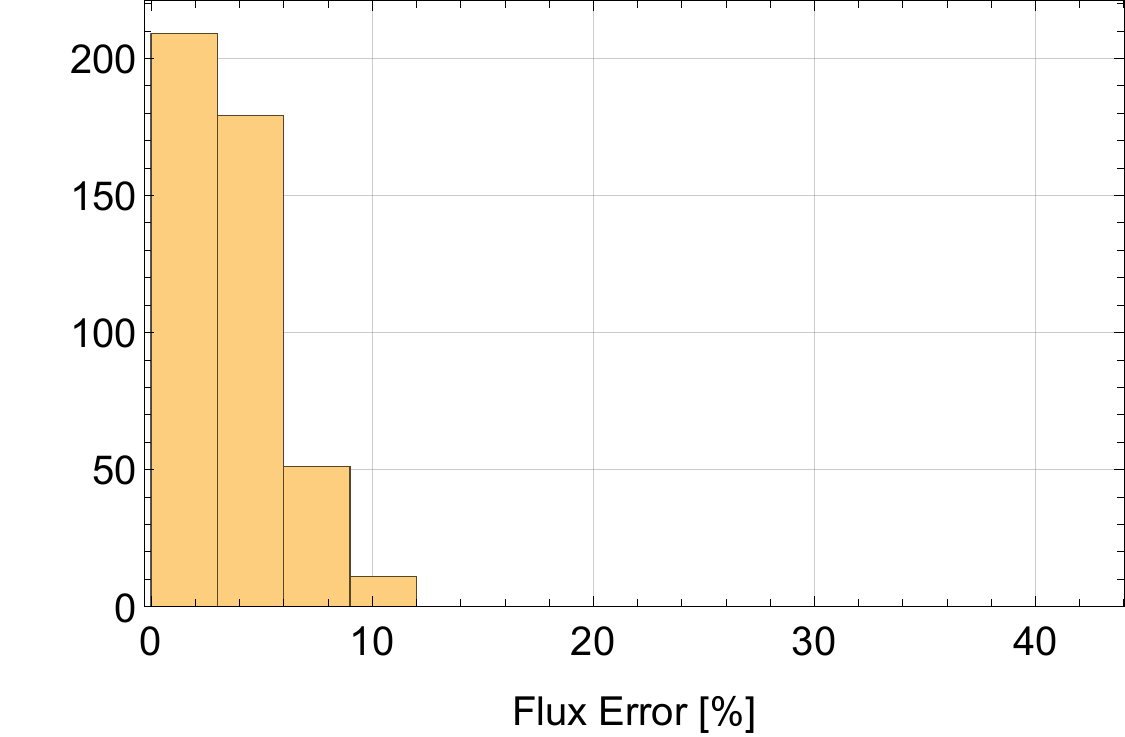}
         \caption{TD truncation of true MM}
         \label{fig:histogramTrunc}
    \end{subfigure}
    \caption{Histograms of flux error, as defined in Eq.~(\ref{eq:fluxError}), of (a) the extrapolated MMs based on assuming the dominant process is Fresnel reflection and (b) the nearest TD truncations of the true MMs relative to the true MMs. Each histogram contains the data over all textures, all geometries, and both wavelengths. The narrower distribution in (b) as compared to (a) indicates that the assumption of a TD eigenspectrum is not the largest source of error. The mean of (a) is 11.06\% and the mode is 1.03\%. The mean of (b) is 3.61\% and the mode is 0.54\%.}\label{fig:histograms}
\end{figure}

To compare MMs with a single-valued metric, the flux error $\epsilon$ is defined as
\begin{equation}
    \epsilon=\frac{\sum_j^{40}|p_j-\widetilde{p}_j|}{\sum_j^{40}p_j},
    \label{eq:fluxError}
\end{equation}
where $p_j$ are the elements of the flux vector simulated by applying the RGB950 measurement matrix $\mathbf{W}_{40}$ to the ground truth MM (\emph{i.e.} the full reconstruction) and $\widetilde{p}_j$ are the elements of the flux vector simulated by applying $\mathbf{W}_{40}$ to the MM being tested. This can be interpreted as adding up all the discrepancies and normalizing by the total expected flux. A flux error $\epsilon=0$ would mean that the two MMs yield the same RGB950 measurements. This physical interpretation is the motivation for choosing the Eq.~\ref{eq:fluxError} as our figure of merit instead of a sum of squared differences between two MMs. Furthermore, small disagreements in multiple off-diagonal MM elements could yield a small squared difference in MM elements, but be an appreciable retardance difference.

Figure~(\ref{fig:histograms}) shows histograms of flux errors calculated from the same flux vectors as in Vid.~(\ref{vid:fluxVectors}) but also includes the other textures. Figure~(\ref{fig:histogramExtrap}) is the histogram of flux errors between the full reconstruction MMs and the MMs extrapolated from linear Stokes images using an assumed Fresnel reflection dominant process. The sources of error are measurement noise, the assumed dominant process, and the assumption of a TD eigenspectrum. The mean is 11.65\% and the mode is 1.03\%. Figure~(\ref{fig:histogramTrunc}) is the histogram of flux errors between the full reconstruction MMs and those same MMs truncated to have a TD eigenspectrum. The process of TD truncation, explained above, preserves the exact dominant process and is not a new noise-realization, so the only source of error is the difference in eigenspectrum. The mean is 3.61\% and the mode is 0.54\%. The narrower distribution in Fig.~(\ref{fig:histogramTrunc}) as compared to Fig.~(\ref{fig:histogramExtrap}) indicates that the assumption of a TD eigenspectrum is not the largest source of error.

\begin{figure}[H]
\centering
    \makebox[5pt]{{\includegraphics[width=0.6\columnwidth]{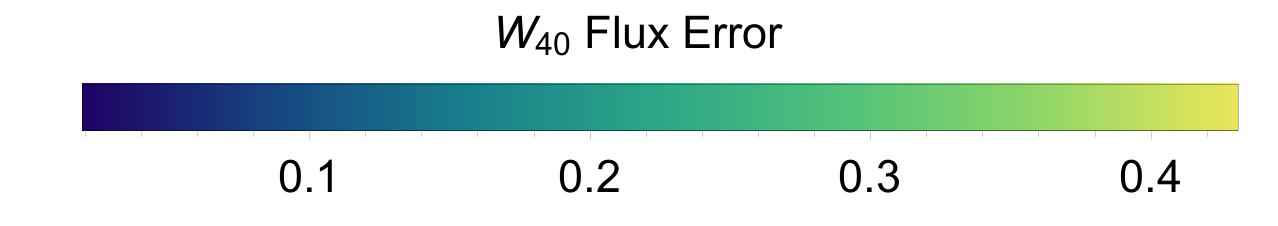}}}\\
    \centering
    \begin{subfigure}[b]{0.49\textwidth}
         \centering
         \includegraphics[width=\textwidth]{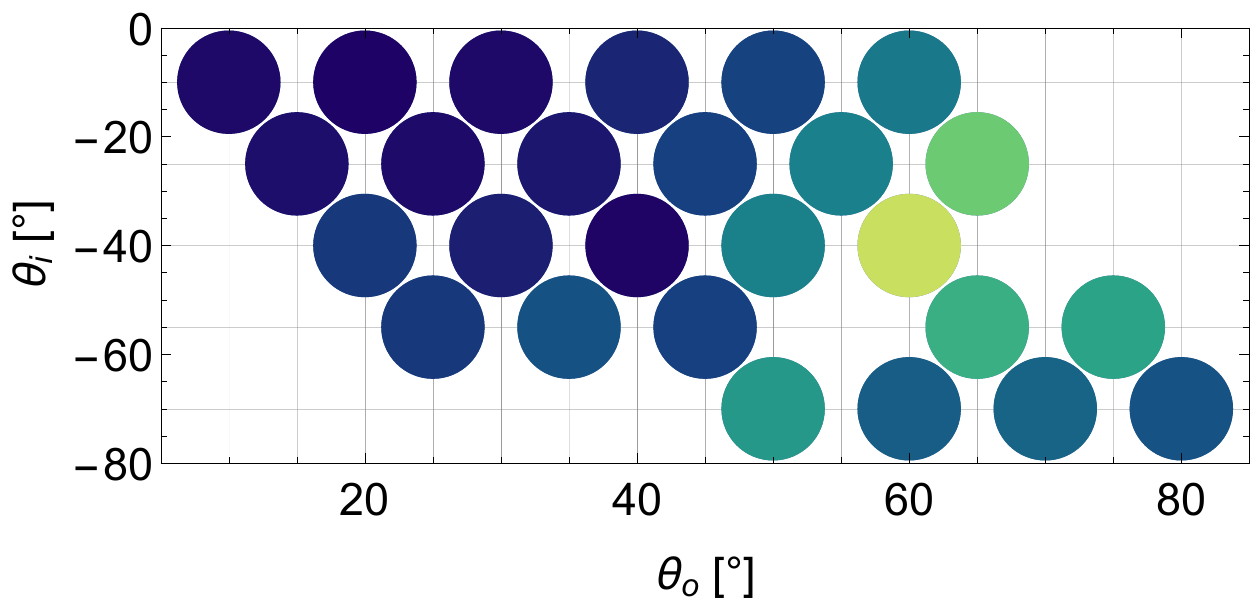}
         \caption{Smoothest brick, 451 nm}
         \label{fig:smooth451}
    \end{subfigure}
         \hfill     
    \begin{subfigure}[b]{0.49\textwidth}
         \centering
         \includegraphics[width=\textwidth]{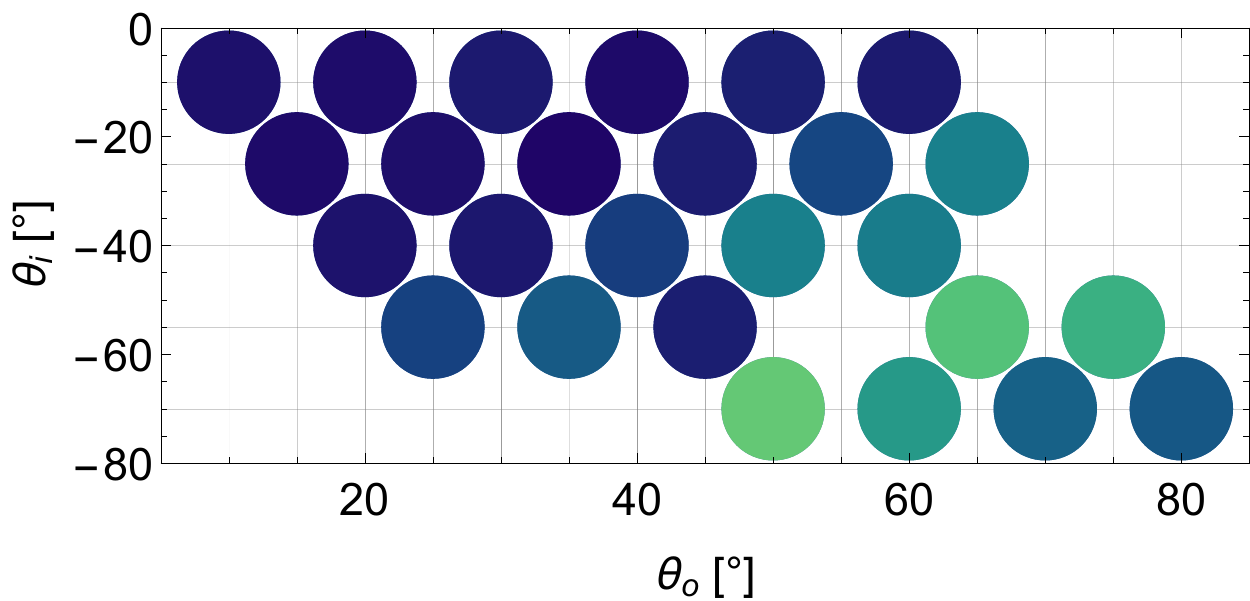}
         \caption{Roughest brick, 451 nm}
         \label{fig:rough451}
    \end{subfigure}
    \centering
   \centering
    \begin{subfigure}[b]{0.49\textwidth}
         \centering
         \includegraphics[width=\textwidth]{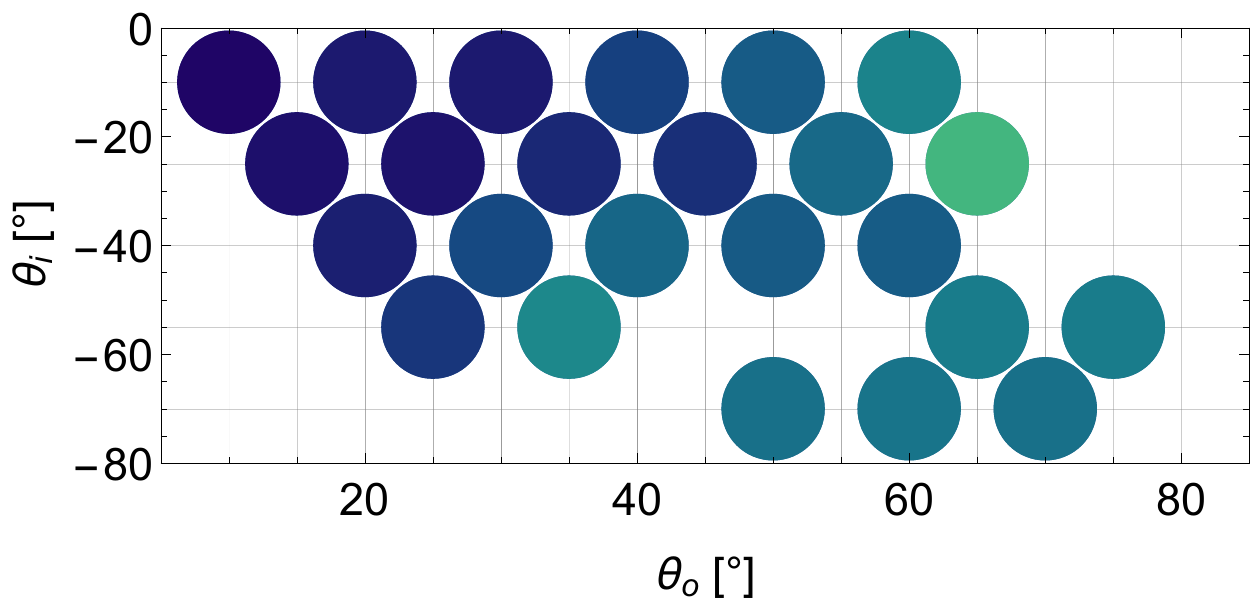}
         \caption{Smoothest brick, 662 nm}
         \label{fig:smooth662}
    \end{subfigure}
         \hfill     
    \begin{subfigure}[b]{0.49\textwidth}
         \centering
         \includegraphics[width=\textwidth]{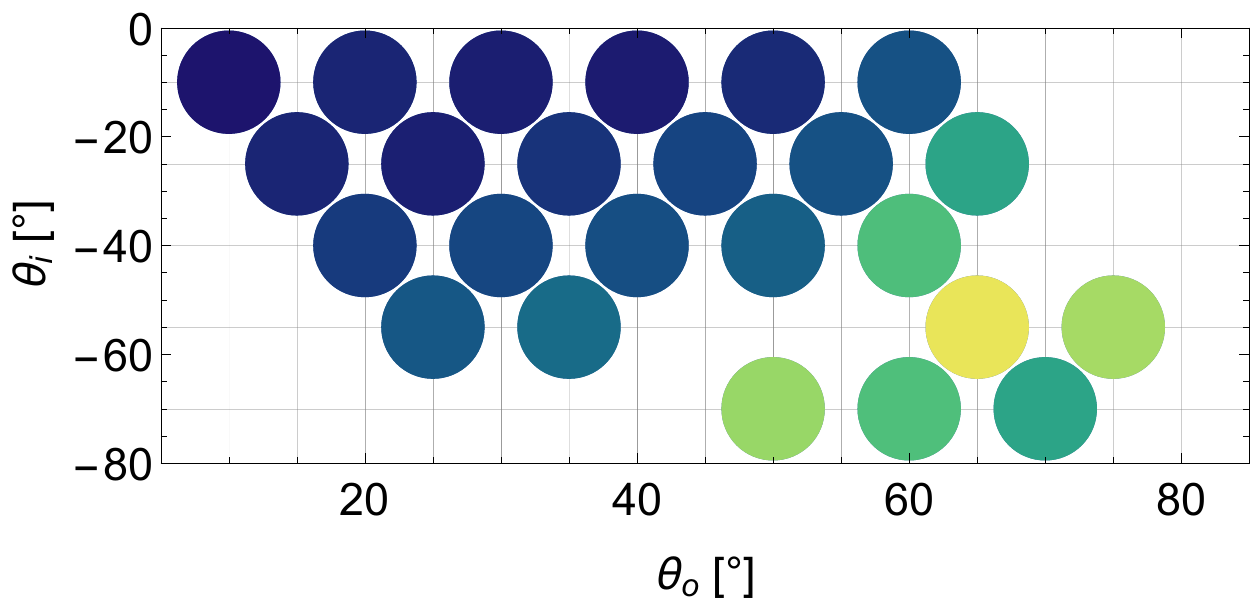}
         \caption{Roughest brick, 662 nm}
         \label{fig:rough662}
    \end{subfigure}
    \caption{Flux vector error $\epsilon$, as defined in Eq.~(\ref{eq:fluxError}), of the LEGO brick MMs extrapolated from linear Stokes images and MMs reconstructed by the RGB950 plotted versus acquisition geometry for (a,c) the smoothest brick and (b,e) the roughest brick. (a) and (b) are the high albedo case and (c) and (d) are the low albedo case. Each flux vector was calculated by averaging the normalized MM on a $2\times2$ pixel region then applying $\mathbf{W}_{40}$ to simulate what the RGB950 would measure. Each point corresponds to an acquisition geometry according to Table~\ref{tab:geom}. Geometries at which the dynamic range of the linear Stokes camera caused non-physical MM extrapolations are omitted. }\label{fig:fluxError}
\end{figure}

For both bricks and both wavelengths, Fig.~(\ref{fig:fluxError}) shows that the larger flux errors tend to occur for the larger incident and scattering angles. The maximum flux error is 0.42 which occurs for the rough brick under 662 nm illumination at $\theta_i=-60^\circ$, $\theta_o=65^\circ$, despite the maximum error in $\xi_0$ occurring for the smooth brick at 451 nm at $\theta_i=-40^\circ$, $\theta_o=60^\circ$. 

\begin{table}[H]
\caption{The flux error $\epsilon$, as defined in Eq.~(\ref{eq:fluxError}) averaged over acquisition geometry for each brick. Acquisition geometries that produced non-physical MM extrapolations due to the dynamic range of the linear Stokes camera are omitted.} 
\label{tab:error}
\begin{center}       
\begin{tabular}{|l||l|l||l|l|l|l|l|l|l|}
\hline
\rule[-1ex]{0pt}{3.5ex}  Brick Ra [microns] & 0.35 & 0.49 & 0.56 & 1.26 & 1.68 & 2.62 & 3.35 & 3.55 & 6.32 \\
\hline\hline
\rule[-1ex]{0pt}{3.5ex}  $\epsilon$ at 451 nm [\%] & 11.50 & 11.40 & 8.96 & 8.33 & 7.79 & 11.06 & 13.28 & 12.28 & 9.93  \\
\hline
\rule[-1ex]{0pt}{3.5ex}  $\epsilon$ at 662 nm [\%]  & 10.42 & 8.42 & 7.23 & 12.82 & 8.75 & 12.95 & 15.55 & 15.16 & 14.89 \\
\hline 
\end{tabular}
\end{center}
\end{table}

Table~\ref{tab:error} reports the flux error $\epsilon$ averaged over acquisition geometry for each brick and at both wavelengths. Both the overall minimum and maximum flux errors occur for 662 nm illumination on the 0.56 micron and 3.35 micron Ra bricks, respectively. Since these extrema do not correspond to the smoothest or roughest textures, it is likely that texture is not the dominant contributing factor to the error. Averaging over texture, the high-albedo case has an error of 10.50\% and the low-albedo case has an error of 11.65\%. 

\section{Conclusion}
For materials described by the simple triple-degenerate polarized light scattering model, this work proposes a new and simplified way to measure the Mueller matrix. While typical Mueller polarimetry requires 16 or more polarimetric measurements to reconstruct a MM, the TD MM model allows for extrapolation from as few as 2 measurements when the dominant process is known \emph{a priori}. Existing methods for extrapolating full MMs from partial polarimetry require non-depolarizing MMs\cite{Ossikovski_9element,SWAMI201318} or, at most, a rank-two coherency matrix\cite{Arteaga_12element}. This work is the only method known to the authors for extrapolating full-rank depolarizing MMs. Additionally, this method is compatible with existing DoFP polarimeter technology and therefore can be made into a snapshot polarimeter.

To demonstrate the method, extrapolations at different geometries of LEGO bricks with varying roughness are performed with a commercial linear Stokes camera and compared to the full MM polarimeter reconstructions. The depolarization, which varies with surface roughness, is apparent even on visual inspection of the diagonal elements of the extrapolated MMs. 
Over varying texture, geometry, and albedo, the partial polarimetric extrapolations achieve flux error mean and mode of 11.06\% and 1.03\%, respectively, despite a 10$\times$ reduction in the number of polarimetric measurements.

\appendix    

\section{Calculation of Rotated Fresnel Reflection Matrix}
\label{sect:fresnel}
From the vectors $\widehat{\boldsymbol{\omega}}_i$ and $\widehat{\boldsymbol{\omega}}_o$, the Fresnel reflection matrix at a given scattering geometry can be found using conventions from polarization ray trace calculus then converting to a MM formalism. 
\subsection{Unrotated Fresnel Reflection}
Based on the microfacet assumption, there are subresolution facets which cause specular reflection from $\widehat{\boldsymbol{\omega}}_i$ to $\widehat{\boldsymbol{\omega}}_o$. The surface normal of such a microfacet is called the halfway vector $\widehat{\mathbf{h}}$ and is calculated by
\begin{equation}
    \widehat{\mathbf{h}}=\frac{\widehat{\boldsymbol{\omega}}_o-\widehat{\boldsymbol{\omega}}_i}{|\widehat{\boldsymbol{\omega}}_o-\widehat{\boldsymbol{\omega}}_i|}.
\end{equation}
The angle of incidence onto the facet is called the difference angle $\theta_d$ and is calculated by
\begin{equation}
    \theta_d=\arccos(\widehat{\mathbf{h}}\cdot(-\widehat{\boldsymbol{\omega}}_i)).
\end{equation}
From the angle of incidence, the Fresnel reflection coefficients are
\begin{eqnarray}
    r_s(\theta_d)=\frac{\cos(\theta_d)-n\sqrt{1-\frac{sin^2(\theta_d)}{n^2}}}{\cos(\theta_d)+n\sqrt{1-\frac{sin^2(\theta_d)}{n^2}}} & \text{and} & r_p(\theta_d)=\frac{n\cos(\theta_d)-\sqrt{1-\frac{sin^2(\theta_d)}{n^2}}}{n\cos(\theta_d)+\sqrt{1-\frac{sin^2(\theta_d)}{n^2}}},
\end{eqnarray}
where $n$ is an estimate of the index of refraction for the measured material and the index of the incident material is assumed to be 1. The Mueller matrix for Fresnel reflection is given in Eq.~(\ref{eq:Fresnel}).


\subsection{Coordinate System of Microfacet}
The input eigenpolarization basis $\widehat{\boldsymbol{\Sigma}}_i$-$\widehat{\boldsymbol{\Pi}}_i$ for the microfacet with normal $\widehat{\mathbf{h}}$ in the transverse plane to $\widehat{\boldsymbol{\omega}}_i$ is calculated by
\begin{eqnarray}
    \widehat{\mathbf{\Sigma}}_i=\widehat{\mathbf{h}}\times\widehat{\boldsymbol\omega}_i & \text{and} & \widehat{\mathbf{\Pi}}_i=\widehat{\boldsymbol\omega}_i\times\widehat{\mathbf{\Sigma}}_i.
\end{eqnarray}
The output eigenpolarization basis $\widehat{\boldsymbol{\Sigma}}_o$-$\widehat{\boldsymbol{\Pi}}_o$ for the microfacet with normal $\widehat{\mathbf{h}}$ in plane transverse to $\widehat{\boldsymbol{\omega}}_o$ is calculated by
\begin{eqnarray}
    \widehat{\mathbf{\Sigma}}_o=\widehat{\mathbf{h}}\times\widehat{\boldsymbol\omega}_o & \text{and} & \widehat{\mathbf{\Pi}}_o=\widehat{\boldsymbol\omega}_o\times\widehat{\mathbf{\Sigma}}_o.
\end{eqnarray}
Greek letters are used to differentiate the microfacet s-p basis coordinates from the macrosurface s-p basis coordinates conventionally used in polarization ray tracing.

\subsection{Coordinate Systems of PSG and PSA}
When light is reflected off a surface, the polarization states also undergo a geometric transform independent of the material properties. The polarization state is initially parameterized in the coordinate system of the PSG and ends in the coordinate system of the PSA. Using a point source illumination model and a pinhole camera model, the polarization states are best described using a double pole basis.

Local basis vectors in a double-pole system are determined by an antipole direction, a rotation matrix from the antipole to the propagation direction, and a reference basis vector.
With the object centered at the origin, the antipole directions of the input and output, $\widehat{\mathbf{a}}_{i}$ and $\widehat{\mathbf{a}}_{o}$, respectively, are calculated by
\begin{eqnarray}
    \widehat{\mathbf{a}}_{i}=-\frac{\mathbf{s}}{|\mathbf{s}|} & \text{and} & \widehat{\mathbf{a}}_{o}=\frac{\mathbf{c}}{|\mathbf{c}|}
\end{eqnarray}
where $\mathbf{s}$ is the coordinate of the point source and $\mathbf{c}$ is the coordinate of the camera pinhole. A 3$\times$3 rotation matrix by angle $\phi$ about the axis $\widehat{\mathbf{r}}=(r_x,r_y,r_z)$ is calculated by
\begin{equation}
    \mathbf{R}_3(\widehat{\mathbf{r}},\phi)=\begin{bmatrix}
    r_x^2(1-\cos\phi)+\cos\phi & r_xr_y(1-\cos\phi)-r_z\sin\phi & r_xr_z(1-\cos\phi)+r_y\sin\phi \\
    r_yr_x(1-\cos\phi)+r_z\sin\phi & r_y^2(1-\cos\phi)+\cos\phi & r_yr_z(1-\cos\phi)-r_x\sin\phi\\
    r_zr_x(1-\cos\phi)-r_y\sin\phi & r_zr_y(1-\cos\phi)+r_x\sin\phi & r_z^2(1-\cos\phi)+\cos\phi.
\end{bmatrix}
\end{equation}
The basis vectors for the input direction $\widehat{\boldsymbol\omega}_i$ are calculated using rotation axis $\widehat{\mathbf{r}}_i$ and rotation angle $\phi_i$
\begin{eqnarray}
        \widehat{\mathbf{r}}_{i}=\widehat{\boldsymbol\omega}_i\times \widehat{\mathbf{a}}_{i} & \text{and} &     \phi_{i}=-\arccos(\widehat{\boldsymbol\omega}_i\cdot \widehat{\mathbf{a}}_{i}),
\end{eqnarray}
basis vectors for the outward direction $\widehat{\boldsymbol\omega}_o$ are calculated using rotation axis $\widehat{\mathbf{r}}_o$ and rotation angle $\phi_o$
\begin{eqnarray}
            \widehat{\mathbf{r}}_{o}=\widehat{\boldsymbol\omega}_o\times \widehat{\mathbf{a}}_{o}
& \text{and} &     \phi_{o}=-\arccos(\widehat{\boldsymbol\omega}_o\cdot \widehat{\mathbf{a}}_{o}).
\end{eqnarray}
Assuming the global vertical is $\widehat{\mathbf{y}}=(0,1,0)$, then the local vertical polarization in the PSG and PSA coordinate systems, $\widehat{\mathbf{y}}_{i,loc}$ and $\widehat{\mathbf{y}}_{o,loc}$ respectively, are calculated by
\begin{eqnarray}
                \widehat{\mathbf{y}}_{i,loc}=\mathbf{R}_3(\widehat{\mathbf{r}}_{i},\theta_i)\widehat{\mathbf{y}} & \text{and} &     \widehat{\mathbf{y}}_{o,loc}=\mathbf{R}_3(\widehat{\mathbf{r}}_{o},\theta_o)\widehat{\mathbf{y}}.
\end{eqnarray}
\subsection{Rotated Fresnel Mueller Matrix}
The rotation matrix in Mueller calculus has the form
\begin{equation}
    \mathbf{R}(\alpha)=\begin{bmatrix}
    1 & 0 & 0 & 0\\
    0 & \cos2\alpha & -\sin2\alpha & 0\\
    0 & \sin2\alpha & \cos2\alpha & 0\\
    0 & 0 & 0 & 1
    \end{bmatrix}.
\end{equation}
The angle of rotation in the transverse plane $\alpha_i$ from the PSG coordinates to the microfacet basis is calculated by
\begin{equation}
    \alpha_i=2\arccos(\widehat{\mathbf{y}}_{i,loc}\cdot\widehat{\boldsymbol\Pi}_i)\mathrm{sign}((\widehat{\mathbf{y}}_{i,loc}\times\widehat{\mathbf{\Pi}}_i)\cdot\widehat{\mathbf{\omega}}_i)
\end{equation}
and the angle of rotation in the transverse plane from the microfacet basis to the PSA coordinates is calculated by
\begin{equation}
    \alpha_o=2\arccos(\widehat{\boldsymbol\Pi}_o\cdot\widehat{\mathbf{y}}_{o,loc})\mathrm{sign}((\widehat{\mathbf{\Pi}}_o\times\widehat{\mathbf{y}}_{o,loc})\cdot\widehat{\mathbf{\omega}}_o).
\end{equation}
The rotated Fresnel reflection Mueller matrix is then calculated by
\begin{equation}
    \mathbf{M}=\mathbf{R}(-\alpha_o)\mathbf{F}^R(\theta_d)\mathbf{R}(-\alpha_i).
\end{equation}

\subsection*{Disclosures}
The authors declare that they have no conflicts of interest.


\subsection* {Code, Data, and Materials Availability} 
Mueller matrix measurement data from the RGB950 and linear Stokes images from the Sony Triton 5.0MP Polarization Camera can be found at https://doi.org/10.25422/azu.data.19763086


\bibliography{report}   
\bibliographystyle{spiejour}   


\vspace{2ex}\noindent\textbf{Quinn Jarecki} is a third year PhD candidate at the Wyant College of Optical Sciences at the University of Arizona. He received his BS in optical science and BM in music performance from the University of Arizona in 2019. His research interests include full and partial Mueller polarimetric imaging and polarized light scattering.

\vspace{2ex}\noindent\textbf{Meredith Kupinski} is an Assistant Professor of the Wyant College of Optical Sciences at the University of Arizona. She received her MS and PhD degrees in optical science from University of Arizona in 2003 and 2008, respectively. Her research interests include task-relevant metrics for imaging system design, estimation/detection theory, and stochastic system analysis and information quantification. 

\listoffigures
\listoftables

\end{spacing}
\end{document}